\documentclass[12pt,reqno]{amsart}
\usepackage{amsmath,amsfonts,amssymb,amscd,amsthm,amsbsy,epsf}
\usepackage[dvips]{graphicx}
\usepackage[usenames]{xcolor}
\usepackage{comment}

\newtheorem{theorem}{Theorem}

\newtheorem{remark}[theorem]{Remark}
\newtheorem{definition}[theorem]{Definition}

\newcommand\beq[1]{ \begin{equation}\label{#1} }
\newcommand{\eeq}{ \end{equation} }

\newcommand\beqa[1]{ \begin{eqnarray} \label{#1}}
\newcommand{\eeqa}{ \end{eqnarray} }
\newcommand{\beqano}{ \begin{eqnarray*} }
\newcommand{\eeqano}{ \end{eqnarray*} }
\newcommand\equ[1]{{\rm (\ref{#1})}}

\def\ep{\varepsilon}

\def\integer{{\mathbb Z}}
\def\nat{{\mathbb N}}

\def\real{{\mathbb R}}
\def\torus{{\mathbb T}}

\begin{document}

\title[Spin-orbit secondary resonances]
{Accurate modelling of the low-order secondary resonances 
in the spin-orbit problem}

\author[I. Gkolias]{Ioannis Gkolias}
\address{
Department of Aerospace Science and Technology,
Politecnico di Milano, Via la Masa 34, 20156 Milan, Italy}
\email{gkolias@mat.uniroma2.it}

\author[C. Efthymiopoulos]{Christos Efthymiopoulos}
\address{
Research Center for Astronomy, Academy of Athens, 
Soranou Efessiou 4, 11527 Athens, Greece}
\email{cefthim@academyofathens.gr}

\author[A. Celletti]{Alessandra Celletti}
\address{
Department of Mathematics, University of Rome Tor Vergata, 
Via della Ricerca Scientifica 1, 00133 Rome, Italy}
\email{celletti@mat.uniroma2.it}

\author[G. Pucacco]{Giuseppe Pucacco}
\address{
Department of Physics, University of Rome Tor Vergata, 
Via della Ricerca Scientifica 1, 00133 Rome, Italy}
\email{pucacco@roma2.infn.it}

\maketitle

\baselineskip=18pt              

\begin{abstract}
We provide an analytical approximation to the dynamics in each of 
the three most important low order secondary resonances (1:1, 2:1, 
and 3:1) bifurcating from the synchronous primary resonance in the 
gravitational spin-orbit problem. To this end we extend the perturbative 
approach introduced in \cite{MNRASpaper}, based on normal form series 
computations. This allows to recover analytically all non-trivial 
features of the phase space topology and bifurcations associated 
with these resonances. Applications include the characterization 
of spin states of irregular planetary satellites or double systems 
of minor bodies with irregular shapes. The key ingredients of our 
method are: i) the use of a \emph{detuning} parameter measuring the 
distance from the exact resonance, and ii) an efficient scheme to 
\emph{`book-keep'} the series terms, which allows to simultaneously 
treat all small parameters entering the problem. Explicit formulas 
are provided for each secondary resonance, yielding i) the time 
evolution of the spin state, ii) the form of phase portraits, 
iii) initial conditions and stability for periodic solutions, 
and iv) bifurcation diagrams associated with the periodic orbits. 
We give also error estimates of the method, based on analyzing 
the asymptotic behavior of the remainder of the normal form series. 
\end{abstract}

\maketitle

\vglue.1cm

\noindent \bf Keywords. \rm Normal form -- primary and secondary resonances -- spin--orbit problem.

\vglue.1cm

\section{Introduction}
\label{sec:intro}

The study of resonant configurations is of primary importance in many 
astronomical problems. One of the most frequently observed commensurabilities 
in our Solar system is that between the orbital and the rotational period 
of natural satellites. Our Moon, for example, is locked in a synchronous 
(1:1) spin-orbit resonance and this is probably the case also for all large planetary 
satellites. In a simple spin-orbit coupling model, the dynamics about the 
synchronous resonance can be described with a pendulum approximation. The 
phase-space is separated by a separatrix into a rotation and a libration 
domain. The frequency of the librations is determined to a first-order 
approximation by the shape of the satellite. For particular values of the 
\emph{asphericity} parameter, used to measure the divergence of the real 
shape from a sphere, this frequency can become resonant with the orbital frequency. 
This situation, known as a {\it secondary resonance}, creates a non-trivial 
topology in the synchronous resonance librational domain which has to be 
studied further.

In astronomical literature, examples of the study of secondary resonances 
around the synchronous primary resonance are motivated by possible connections 
to the problem of tidal evolution of systems such as a satellite with 
aspherical shape around a planet, or a double configuration of minor bodies 
(e.g. asteroids) where one or both bodies have irregular shapes. An example 
of the former case is Enceladus: it was originally conjectured (\cite{Wisdom2004}) 
that the asphericity ratio of this satellite would make possible a past 
temporary trapping into the 3:1 secondary resonance located within the 
synchronous spin-orbit resonance with Saturn. Such a scenario would 
justify an amount of tidal heating substantially larger than far from 
the secondary resonance. The efficiency of this scenario
was questioned as Cassini's observations reduced Enceladus' 
estimated asphericity closer to $\ep\approx 1/4$ rather than $1/3$ 
(\cite{poretal2006}; see the review by \cite{meywis2007}). On the other 
hand, the overall role that secondary resonances could have played for 
the tidal evolution of planetary satellites towards their final synchronous 
state is a largely open problem. As regards minor planetary satellites or 
double minor bodies (e.g. double asteroids), exploration of the subject is 
still bounded by the scarcity of observations (see e.g. \cite{boh}). A question 
of central interest regards predicting changes in the {\it stability} character 
of a certain spin `mode' (or periodic orbit) associated with a resonance, 
as the main parameters of the problem (eccentricity, asphericity) are 
varied. Varying the parameters leads to bifurcations of new periodic 
orbits, accompanied by a change of stability of their parent orbits. 
For secondary resonances $l:k$ of order $|l|+|k|> 4$, such bifurcations 
are described by a general theory (see, for example, \cite{arn1978}).
Instead, for low order resonances ($2\leq |l|+|k|\leq 4$) such transitions 
are case-dependent, and they lead to important changes in the topology 
of the phase portrait in the neighbourhood of one resonance. Besides 
theoretical interest in modelling these cases, the determination of 
stability of the various resonant modes can be useful to the interpretation 
of observations. An additional motivation stems from the need for precise  
models of spin-orbit motion in connection with future planned missions 
to double minor body systems. 

With the above applications in mind, in the present paper we discuss 
the implementation of our method recently introduced in \cite{MNRASpaper} 
with the aim to provide an {\it analytical modelling} allowing to fully 
reproduce the dynamics of the 3:1, 2:1 and 
1:1 secondary resonances around the synchronous primary spin orbit resonance. 
Besides demonstrating the ability to analytically deal with all peculiarities 
encountered in the phase space features and bifurcation properties of these 
secondary resonances, the provision of analytical formulas with high precision 
is of practical utility, as it can substitute expensive numerical treatments 
with practically no loss of accuracy. In fact, we make an analysis of 
the error introduced in our approximation, based on well known methods 
used in asymptotic analysis of series expansions in classical perturbation 
theory. More precisely, after computing a Hamiltonian normal form for 
the secondary resonance, we measure the goodness of the approximation 
by the estimate of the {\it remainder function}, whose size is determined 
by two principal factors: i) the way we `book-keep' the series terms including 
the {\it detuning} as a small parameter in the series (see section 2 
below), and ii) the accumulation of small divisors in the series terms. 
The typical behavior of the size of the remainder is that it decreases 
up to a certain order and then it increases. The order at which the size 
of the remainder attains its minimum is called the \sl optimal order \rm 
of the normal form. In this work we outline a procedure to estimate the 
optimal order, and hence obtain explicit estimates of the error of our 
analytical approximation. In fact, a key result is that our normal form 
construction, albeit non-standard in the way we `book-keep' the Hamiltonian 
terms, still exhibits the desired asymptotic behavior of 
more conventional constructions, as, e.g., multivariate series in powers 
of more than one small parameters (see for example \cite{Sans2014}). 

The paper is organized as follows. The general problem is introduced in 
section~\ref{sec:H}. The normalization process is discussed in a general 
setting in section~\ref{sec:gp}, along with a demonstration of how 
estimates of the errors follow from an asymptotic analysis of the 
normal form's remainder. The specific application to the description 
of secondary resonances in the spin-orbit problem is given in 
section~\ref{sec:model}, with concrete applications to the 1:1, 
2:1 and 3:1 secondary resonances. Error analysis for each secondary 
resonance is discussed in section~\ref{sec:asym}. Our results are 
summarised in section~\ref{sec:conclusions}. Explicit 
formulas for use in analytic computations are provided 
in the Appendix.

\section{Hamiltonian of the spin-orbit problem}\label{sec:H}
The Hamiltonian describing the orbital and rotational coupling of a 
satellite in a Keplerian orbit, rotating about one of its primary axes of 
inertia, which is assumed perpendicular to the orbital plane, is given by \cite{goldpeale,MD,zamp1}:
\beq{eq:fullham}
H(p_{\theta},\theta,t)=\frac{p_{\theta}^2}{2}-\nu^2\frac{\ep^2}{4} 
\frac{a^3}{r^3(t)}\cos(2 \theta-2f(t))\ ,
\eeq
where $\theta$ is the angle formed by the largest physical axis of the satellite 
and the orbit apsis line, $a$ is the orbit's semi-major axis, $\nu$ is the orbital
frequency, $f$ the true anomaly, $r$ the distance between the two bodies and 
$\ep$ the \emph{asphericity} parameter defined as:
\beq{eq:asphericity}
\ep = \sqrt{\frac{3 (B-A)}{C}},
\eeq
where $A,B,C$ are the moments of inertia of the satellite ($C$ is the one
corresponding to the rotational axis), and we assume $A \leq B \leq C$ (see \cite{Lhotka2013} for a discretized version, \cite{Cel2014} for a dissipative version of the spin-orbit equation). 

We choose units such that $a = \nu = 1$. Both the true anomaly $f=f(t)$ 
and the orbital radius  $r=r(t)$ are known functions of the time and 
can be expanded in Fourier series. Therefore, making explicit the time
dependence, the spin-orbit Hamiltonian takes the form:
\begin{equation}
H(p_{\theta},\theta,t)=\frac{p_{\theta}^2}{2}-\frac{\ep^2}{4} 
\sum_{m \neq 0,m=-\infty}^{m=\infty} W\left(\frac{m}{2},e\right) 
\cos(2 \theta- m t )\ ,
\label{eq:fullhamfour}
\end{equation}
where the coefficients $W=W\left(\frac{m}{2},e\right)$ are the classical 
$G$ functions of \cite{Kau1966} and they are series in the eccentricity 
of order $e^{|m-2|}$ (\cite{Cay1861}):
\beqano
W\left(\frac12,e\right)&=&-\frac{e}{2}+\frac{e^3}{16}+\mathcal{O}(e^5),\nonumber\\
W(1,e)&=&1-\frac{5}{2}e^2+\frac{13}{16}e^4+\mathcal{O}(e^6),\nonumber\\
W\left(\frac32,e\right)&=&\frac{7}{2}e-\frac{123}{16}e^3+\mathcal{O}(e^5)
\eeqano

We consider now an extended phase-space by introducing a \sl dummy \rm action 
$p_2$, conjugated to the time variable with frequency equal to the orbital 
frequency (which is equal to 1). The extended Hamiltonian reads as
$$
H (p_1,p_2,q_1,q_2)=\frac{p_1^2}{2}+p_2-\frac{\ep^2}{4} 
\sum_{m \neq 0,m=-\infty}^{m=\infty} W\left(\frac{m}{2},e\right)
\cos(2 q_1 - m q_2 )\ .
$$
Introducing  the \sl resonant angle \rm
$$
\psi = q_1 - \frac{p}{q} q_2
$$ 
through the canonical transformation
\beq{eq:rotransf1}
 p_1 = p_{\psi} + \frac{p}{q}\, \quad p_2 = p_{\phi} - \frac{p}{q} p_{\psi}\, \quad \psi = q_1 - \frac{p}{q} q_2\, \quad \phi = q_2 
\eeq
for some $p$, $q$ integers, the Hamiltonian takes the form
\begin{equation}
H=p_{\phi}+\frac{p^2_{\psi}}{2}-\frac{\ep^2}{4} 
\sum_{m \neq 0,m=-\infty}^{m=\infty} W\left(\frac{m}{2},e\right)
\cos(2 \psi + \left(2 (p/q) - m\right) \phi ).
\label{HR}
\end{equation}
The ratio $p/q$ in \equ{eq:rotransf1} is chosen according to which primary 
resonance we are interested in studying. For $m=2 (p/q)$ the angle 
$\phi$ vanishes from the arguments of the trigonometric terms and the 
Hamiltonian takes the form:
\begin{equation}
H=p_{\phi}+\frac{p^2_{\psi}}{2}-\frac{\ep^2}{4} 
W\left(\frac{p}{q},e\right) \cos(2 \psi)+H_{\textrm{nonres}}.
\label{HR2}
\end{equation}
We remark that the resonant part of the Hamiltonian \equ{HR2} is the 
sum of the dummy action $p_{\phi}$ and a pendulum-like Hamiltonian 
in the resonant angle $\psi$.

For the synchronous (1:1) resonance the transformation \eqref{eq:rotransf1} 
reads as
$$
p_1=p_\psi +1\ , \quad p_2=p_{\phi}-p_{\psi}\ , 
\quad \psi = q_1 - q_2\ , \quad \phi = q_2\ ,
$$
and the resonant part of the Hamiltonian is
\begin{equation}
H_\textrm{res}=p_{\phi}+\frac{p^2_{\psi}}{2}-\frac{\ep^2}{4} W(1,e) 
\cos(2 \psi),
\label{HR3}
\end{equation}
where $W(1,e) = 1-\frac{5}{2}e^2+\frac{13}{16}e^4+\ldots$. To describe 
the librations around the primary resonance, through the 
Taylor series $\cos(2 \psi) = 1-2 \psi^2 
+ \ldots$, we get the Hamiltonian:
\begin{equation}\label{eq:hamprim11}
H=p_{\phi}+\frac{p^2_{\psi}}{2}+\frac{\ep^2}{2} \psi^2 + 
H_{\text{pert}}(\psi,\phi;e,\ep)\ ,
\end{equation}
where, $H_{\textrm{pert}}$ is polynomial in $\psi$.

The integrable part of the Hamiltonian introduces the  
\sl unperturbed \rm frequencies $\omega_1=1, \omega_2=\ep$. 
Finally, we introduce the action angle variables ($J,u$)
through
\begin{equation}\label{AARV}
\psi = \sqrt{\frac{2 J}{\ep}} \sin{u}\ , 
\quad p_{\psi} = \sqrt{2 J\ep} \cos{u}\ , 
\quad J_{\phi} = p_{\phi}\ ,
\end{equation}
which brings our Hamiltonian into the following form:
\beq{eq:Hpertgeneral}
H=J_{\phi}+ \ep J + H_{\text{pert}}(J,u,\phi;e,\ep)\ .
\eeq
The perturbing part $H_{\text{pert}}$ is a Fourier series in 
$u,\phi$ of the form
$$
H_{\text{pert}}(J,u,\phi;e,\ep) = 
\sum_{k_0,k_1,k_2} c_{k_0k_1k_2} (e,\ep) 
J^{k_0\over 2} {\rm e}^{i (k_1 u + k_2 \phi)}\ , 
\quad k_0,k_1,k_2 \in \mathbb N\ .
$$
\section{General normal form theory}\label{sec:gp}
In this section we discuss our proposed canonical normalization 
procedure and generalise our method for the study of an arbitrary order 
secondary resonance appearing in the vicinity of a primary resonance 
that can be described locally by a pendulum approximation. First we 
assume a Hamiltonian model which has the form \eqref{eq:Hpertgeneral}. Then, 
we introduce the main ingredients that will be used to compute the 
normal form: the introduction of a detuning term, measuring the distance 
from the exact resonance, and the ordering of different terms through 
a book-keeping parameter. Finally, we discuss how to obtain error 
estimates based on an optimal normalisation order in our construction.

\subsection{The Hamiltonian}\label{sec:Hamgen}
Consider a general Hamiltonian system with 2 d.o.f.,
depending on a set of $M$ \sl control parameters \rm $c_{\alpha}$ $\alpha=1,...,M$,
which are associated with the specific nature of the problem. 
Let $(J_1,J_2,\phi_1,\phi_2)$ denote
action-angle variables with $(J_1,J_2)\in\real^2$, $(\phi_1,\phi_2)\in\torus^2$.
We consider a Hamiltonian function of the form
\begin{equation}\label{FEX}
H(J_1,J_2,\phi_1,\phi_2;c_{\alpha}) = \sum_{j_1,j_2,k_1,k_2\in \mathbb Z} 
a_{j_1j_2k_1k_2} (c_{\alpha}) J_1^{j_1/2} J_2^{j_2/2}
{\rm e}^{i (k_1 \phi_1 + k_2 \phi_2)}\ ,
\end{equation}
where $a_{j_1j_2k_1k_2}$ are real coefficients depending on the control 
parameters. According to \cite{mho}, we introduce the following definition.

\begin{definition}
The Hamiltonian \equ{FEX} is said to have the {\rm D'Alembert character}, 
whenever for $j_1,j_2\in \mathbb N$, $k_1,k_2 \in \mathbb Z$, the 
following conditions are satisfied:
\beq{eq:dalconds}
j_a \ge |k_a|  , j_a = |k_a|  \pmod{2}, \quad a=1,2.
\eeq
\end{definition}

As showed in \cite{mho}, the Hamiltonian \equ{FEX} is derived from a 
power series of the form $\sum b_{k_1k_2\ell_1\ell_2}
p_1^{k_1}p_2^{k_2}q_1^{\ell_1}q_2^{\ell_2}$, setting 
$p_k=(2 J_k)^{1/2}\cos\phi_k$, $q_k=(2 J_k)^{1/2}\sin\phi_k$, if and only 
if \equ{FEX} has the D'Alembert character. It is useful to note here
that in the derived power series $p_k$ and $q_k$ appear
only in positive integer powers, i.e. $k_1,k_2,\ell_1,\ell_2 \in \nat$. 

\subsection{Detuning}\label{sec:det}
The class of Hamiltonian systems described by \eqref{FEX} includes 
nearly-integrable systems, provided one identifies an integrable 
part and assuming that the remaining terms are \sl small \rm in 
some sense. A typical example which naturally comes out when 
reducing the system around a given resonance is represented by 
a Hamiltonian function which is linear in the actions. This means 
that \eqref{FEX} should admit linear terms independent of the angles, 
taking the form
\beq{LP}
H_0(J_1,J_2)=\omega_1 J_1 + \omega_2 J_2 \ ,
\eeq
where $\omega_a\in\real$, $a=1,2$, denote the \sl unperturbed \rm 
frequencies associated with oscillations in the $(q_1,p_1)$ and 
$(q_2,p_2)$ planes. We will focus on the case in which
there exists a near (albeit not necessarily exact)
commensurability between the unperturbed frequencies, 
which can be expressed in the form
\beq{detkl}
\frac{\omega_1}{\omega_2} - \frac{k}{\ell} \equiv \delta\ ,
\eeq
where $k, \ell \in \mathbb Z$, and $\delta$ is a small real 
parameter which we refer to as the \sl detuning \rm (see \cite{Vf,MP13,MP14}). 
It is important to notice that, in this generic case, the resonance 
is in principle absent from the unperturbed dynamics, but it can 
appear in the perturbed system, once it is \sl triggered \rm by 
the non-linear, higher-order coupling terms. Low-order nearly-resonant 
ratios (namely those with $|k|+|\ell|\le 4$) deserve particular 
attention, since they generate several interesting phenomena 
which will be examined in the following sections.

Having fixed a given ${k}/{\ell}$ (nearly) resonance as in 
\equ{detkl}, a normalization process can be implemented to transform
the original Hamiltonian \eqref{FEX} into a \sl normal form. \rm
As detailed in Section~\ref{sec:norma}, the standard approach is 
that in which the normal form is constructed by imposing the 
conservation of the linear part \eqref{LP}. A resonant 
${k}/{\ell}$ normal form is more generically set-up under the 
hypothesis that the condition \eqref{detkl} is satisfied with 
$\delta=0$, while the small term proportional to the detuning is 
considered as part of the perturbation.

\subsection{Book-keeping}\label{sec:bk}
The Hamiltonian \eqref{FEX} is a series expansion whose terms 
are characterized by three different small parameter scales: 
they are respectively associated with the action variables 
$J_a$ (giving the amplitude of the motion), (a subset of) 
the coupling parameters $c_{\alpha}$ and the detuning $\delta$. 
Powers of each of these quantities appear in the series 
expansions of the original and transformed Hamiltonians. 
Since the normalization is not a unique process, as different 
strategies can be adopted according to various ordering of 
the terms, it is very useful to use a single parameter,
which is able to deal with all sets of small quantities at the 
same time. According, e.g., to \cite{CHR}, we introduce a \sl 
book-keeping \rm parameter $\lambda$, which determines the 
ordering of the various terms in \eqref{FEX} by means of suitable 
substitution rules. Applying such rules, the decrease in size of 
each term will be naturally related to increasing powers of 
$\lambda$.

The rules for assigning the book-keeping parameter to the 
set of the action coordinates, the small control parameters,
and the detuning are implemented as follows:

1. Scaling of the action variables  $J_a$ is the usual procedure 
to account for the ordering of terms of different powers in the 
amplitude of motion. In view of the role played by the linear 
terms in \eqref{LP} and the fact that in the expansion there 
can be altogether smaller action terms with exponents 
$j_a \le 1$, the natural choice is to perform the following 
scaling for powers of the actions:
\beq{bookJ}
J_a^{j_a} \rightarrow \lambda^{{\rm max}[2 j_a -2,0]} 
J_a^{j_a} \ , \quad a=1,2 \ ,
\eeq
where ${\rm max}[q,0]$ denotes the greatest between the relative 
integer $q$ and zero. This choice reflects a natural scaling of 
the oscillating phase-space variables, which transforms
as half-integer powers of $J_a$.

2. Concerning the coupling parameters, we can simplify the 
discussion by making the assumption that among the 
$c_{\alpha}$, $\alpha=1,...,M$, only one of them is small with 
respect to the others and we call it $c_S$. In the example of the
Hamiltonian \eqref{eq:Hpertgeneral}, the role of the small 
parameter is played by the eccentricity $e$. We decide to rescale 
the small parameter $c_S$ as
\beq{booke}
 c_S \rightarrow \lambda \ c_S \ .
\eeq
We stress that in case of more small parameters (e.g., the 
eccentricity and the inclination) the rescaling can be conveniently 
applied to all small control parameters.

3. The detuning parameter introduced in \eqref{detkl} is assumed 
to be small. Therefore, the natural choice is the substitution
\beq{bookd}
 \delta \rightarrow \lambda \ \delta \ .
 \eeq
We recall that in general the parameter $\delta$ may appear not 
only in the linear part \eqref{LP}, but also within higher-order 
terms.\\

Applying the three rules described before to the Hamiltonian 
\eqref{FEX} and rescaling time according to
$
 t \rightarrow \frac{\omega_2}{\ell} \ t \ ,
$
the \sl book-kept \rm Hamiltonian takes the form
\begin{equation}\label{eq:hambookkept}
H=k J_1+ \ell J_2 + \lambda \ell \delta J_2 + \sum_{i} 
\lambda^{i} H_{i}(J_1,J_2,\phi_1,\phi_2;c_{\alpha},\delta) \ , 
\end{equation}
where $H_{i}$, $i \ge 1$, denote terms of progressively higher 
order in $\lambda$.\\

\begin{remark}\label{remarklambda}
We remark that $\lambda$ is a symbol appearing at all orders of 
the expansions; once the normalization procedure is completed, 
the value of $\lambda$ is set to one, thus losing any quantitative 
meaning. Nevertheless, powers of $\lambda$ allow us to group different 
terms in all expansions according to their corresponding order of smallness.
Moreover, the notation $\mathcal{O}_s$ indicates a series of terms of 
powers $s$ or higher in the book-keeping parameter $\lambda$. 
\end{remark}

\subsection{Canonical normalization}\label{sec:norma}

The normalisation approach implemented on the Hamiltonian
 \eqref{eq:hambookkept} consists of finding a change of variables 
from $(J_1,J_2,\phi_1,\phi_2)$ to a new set of coordinates, such 
that the new Hamiltonian is in resonant normal form up to high orders in the 
book-keeping parameter. The normalization can be achieved through 
different approaches; here we choose to implement the so-called 
Hori-Deprit method (see, e.g., \cite{H1966,D1969,gior}), which is based on 
Lie series transformations.

The method consists in finding a sequence of canonical
transformations close to the identity, so that the initial
coordinates $(J_1,J_2,\phi_1,\phi_2) \equiv
(J_1^{(0)},J_2^{(0)},\phi_1^{(0)},\phi_2^{(0)})$ are successively
transformed as
\beq{eq:transvarlie}
(J_1^{(0)},J_2^{(0)},\phi_1^{(0)},\phi_2^{(0)}) \rightarrow
(J_1^{(1)},J_2^{(1)},\phi_1^{(1)},\phi_2^{(1)}) \rightarrow
(J_1^{(2)},J_2^{(2)},\phi_1^{(2)},\phi_2^{(2)}) \ldots
\eeq
The sequence of transformations are determined in such a way that
the transformed Hamiltonian after $n$ normalization steps $H^{(n)}$ takes the form
\beq{hamnf}
H^{(n)}=Z_0+\lambda Z_{1}+\ldots+\lambda^{n} Z_{n} + \lambda^{n+1} H^{(n)}_{n+1} +
\lambda^{n+2} H^{(n)}_{n+2}+\mathcal{O}_{n+3} ,
\eeq
where $\lambda$ denotes the book-keeping parameter.
We refer to the \sl normal form \rm part of the Hamiltonian
\equ{hamnf} as the function
\beq{nfm}
Z^{(n)}=Z_0+\lambda Z_{1}+\ldots+\lambda^{n} Z_{n}\ ,
\eeq
which depends just on the actions in the
non-resonant case, or on the actions and on suitable
combinations of the angles in the resonant case. The functions $Z_j$ 
are determined recursively, together 
with the generating functions of the Lie canonical transformation
by solving suitable homological equations. With reference to \equ{hamnf}, 
we define the \sl remainder \rm function 
after $n$ normalisation steps as the quantity
\beq{rem}
R^{(n)}=\lambda^{n+1} H^{(n)}_{n+1} + \lambda^{n+2} H^{(n)}_{n+2}+\mathcal{O}_{n+3} .
\eeq
The size of $R^{(n)}$ gives a measure of the difference between 
the true dynamics and that provided by the normal form $Z^{(n)}$, 
thus yielding  the size of the error of the normal form approach 
at the order $n$ (see subsection \ref{sec:so} and Section~\ref{sec:asym}).

Using the Hori-Deprit method, the changes of coordinates 
\eqref{eq:transvarlie}
are determined using a sequence of Lie generating functions.
Normalizing up to the order $n$, we consider the generating functions
$\chi_1$, $\chi_2$, ... , $\chi_n$, such that
\begin{eqnarray}\label{lietra}
~J_1&=&\exp(L_{\chi_{n}})\exp(L_{\chi_{n-1}})
\ldots\exp(L_{\chi_1}) J_1^{(n)} \nonumber \\
~~J_2&=&\exp(L_{\chi_{n}})\exp(L_{\chi_{n-1}})
\ldots\exp(L_{\chi_1}) J_2^{(n)} \\
~\phi_1&=&\exp(L_{\chi_{n}})\exp(L_{\chi_{n-1}})
\ldots\exp(L_{\chi_1}) \phi_1^{(n)}  \nonumber \\
~\phi_2&=&\exp(L_{\chi_{n}})\exp(L_{\chi_{n-1}})
\ldots\exp(L_{\chi_1}) \phi_2^{(n)} ,\nonumber
\end{eqnarray}
where $L_{\chi}$ denotes the Poisson bracket operator, 
$$
L_{\chi} (\cdot) \equiv \left\lbrace \cdot, \chi \right\rbrace
$$
and the exponential is defined as
\beq{exp}
\exp(L_\chi)=\sum_{k=0}^\infty \frac1{k!} L_\chi^k\ .
\eeq
In practice, one needs to retain a finite
number $N$ of terms in \equ{rem}, with $N>n$. $N$ will be referred
to as the \sl truncation order. \rm Upon the transformation of
coordinates \equ{lietra}, the Hamiltonian becomes
\beq{lieham}
H^{(n)}= \exp(L_{\chi_{n}})\exp(L_{\chi_{n-1}})...
\exp(L_{\chi_2})\exp(L_{\chi_1})H^{(0)}\ .
\eeq
We remark that this Hamiltonian is composed of the $n$-th order 
normal form and $N-n$ consecutive terms of the remainder series.

The generating functions $\chi_j$, $j=1,2,\ldots n$ are 
determined recursively by solving, at the $r$-th step of the 
normalization  procedure, the following {\em homological equation}:
\beq{eq:homological}
\left\lbrace Z_0, \chi_{r+1} \right\rbrace  + \lambda^{r+1} h^{(r)}_{r+1} = 0\ .
\eeq
The function $Z_0 = k J_1+ \ell J_2$ is named the 
\sl kernel \rm of the normalization procedure, 
while the function $h^{(r)}_{r+1}$ is composed of all terms of 
$H^{(r)}_{r+1}$, whose Poisson bracket with $Z_0$ is different 
from zero. In this way we obtain the function $Z^{(r+1)}=
H^{(r)}_{r+1}-h^{(r)}_{r+1}$ (see \cite{H1966}, \cite{D1969}, 
\cite{CHR} for further details).

The functions $h^{(r)}_{r+1}$, $r=1,2,\ldots,n$ can be written as 
the Fourier sum
$$
h^{(r)}_{r+1} = \sum_{k_1,k_2 \not\in 
\mathcal{ M } } b^{(r)}_{r+1,(k_1,k_2)} (J_1,J_2) 
{\rm e}^{i (k_1 \phi_1 + k_2 \phi_2)}, 
$$
where 
$$
\mathcal{ M } = \left\lbrace \mathbf{k} 
\equiv (k_1,k_2) : k_1 k - k_2 \ell = 0 \right\rbrace\ 
$$
is the \emph{resonant module}. Thus, the solution of the homological 
equation can be written as
$$
\chi_{r+1} =  \sum_{k_1,k_2 \not\in \mathcal{ M } } 
\frac{b^{(r)}_{r+1,(k_1,k_2)} (J_1,J_2)}{k_1 k - k_2 \ell } 
{\rm e}^{i (k_1 \phi_1 + k_2 \phi_2)} \ .
$$
By implementing the above procedure, at each step we obtain a new
Hamiltonian
$$
H^{(r+1)}=\exp{L_{\chi_{r+1}}} H^{(r)}\ ,
$$
which is normalized up to the order $r+1$:
$$
H^{(r+1)}=Z_0+\lambda Z_{1}+\ldots+\lambda^{r} Z_{r} 
+ \lambda^{r+1} Z_{r+1} + \lambda^{r+2} H^{(r+1)}_{r+2}+\mathcal{O}_{r+3}.
$$
\remark{\label{remarktransr}
The functions $H^{(r)}$ depend on the transformed variables $\phi^{(r)},J^{(r)}$.
For simplicity of notation, we hereafter avoid superscripts in the 
notation of canonical variables, assuming correspondence with the order of 
normalization which is provided whenever needed.
}

In the non-resonant case, the function $Z_0$ depends only on 
$J_1, J_2$, while in the resonant case the normal form depends 
also on the combination of the angles $ \ell \phi_1 - k \phi_2$. 
This leads to introduce in a natural way another set of canonical
variables $(J_F,J_R,\phi_{F},\phi_{R})$ for the resonant
Hamiltonian, defined as
\begin{equation*}\label{Treso}
\phi_1 \rightarrow \phi_{R}+\frac{k}{\ell}\phi_{F},\quad \phi_2 \rightarrow \phi_F,\quad
J_1 \rightarrow J_R ,\quad J_2 \rightarrow J_F - \frac{k}{\ell} J_R
\end{equation*}
where the suffix $F$ stands for \sl fast \rm and $R$ stands for 
\sl resonant, \rm so that 
$$
\ell \phi_{R} = \ell \phi_1 - k \phi_2 \ .
$$
The transformed Hamiltonian normal form becomes:
\begin{equation}\label{eq:Hreso}
Z = \ell J_F  + \lambda \delta (\ell J_F - k J_R) + 
\sum_{j=1}^n \lambda^j Z_j(J_F,J_R,\ell \phi_R;c_{\alpha},\delta) \ .
\end{equation}
The action
$$
J_F = {Z_0 \over \ell} = \frac{k}{\ell} J_1+  J_2
$$
is now a constant of the motion, since its conjugate angle $\phi_{F}$ 
is not present in the Hamiltonian. The problem has finally been reduced 
to one degree of freedom and it is an integrable approximation of the 
original non-integrable system \eqref{eq:hambookkept}.
%

\subsection{Orbits and Phase Portraits}\label{sec:phase}
Among the different applications of the normal form, we start by quoting 
that the Hamiltonian \eqref{nfm} (or \eqref{eq:Hreso}) provides an integrable approximation 
of the original system \eqref{eq:hambookkept}, which is more accurate 
than just retaining the lowest order term. For example, in the resonant 
case by analyzing the reduced function (\ref{eq:Hreso}), one can obtain
valuable information about the original system. The solutions of
the real system are encoded in the level curves of the integral
$Z_0$ or, equivalently, the constant energy curves of the
Hamiltonian (\ref{eq:Hreso}). In fact, by trivially integrating the
orbits of (\ref{eq:Hreso}) and back-transforming to the original
variables via the transformation equations (\ref{lietra}), one
obtains highly precise approximations of the time solutions of the real
system, at least in the domain of regular motions.

\subsection{Analytical approximation of the periodic orbits}\label{sec:periodic}
Periodic solutions of the equations of motion play a very important role. 
Several methods have been developed to compute periodic orbits of Hamiltonian 
systems. Taking advantage of the simplified dynamics of the resonant 
normal form (\ref{eq:Hreso}), an explicit formula for the periodic orbits 
associated with the main resonance can be easily derived.

Such periodic solutions correspond to the equilibrium points of
the reduced normal form \equ{HXY11}. Let us denote by
\beq{equipoint}
J_R=J_0\ ,\qquad \phi_R=\phi_0
\eeq
one of these points and by
$$
\omega_F = \frac{\partial Z}{\partial J_F} \Bigg|_{J_R=J_0,\phi_R=\phi_0}
$$
the \sl fast \rm frequency. Fixing a level set for $J_F$ and using the 
same procedure as for any other phase-space function, we back-transform 
the equilibrium point to a solution in terms of the original variables 
$(J_1,J_2,\phi_1,\phi_2)$:
\begin{equation}\label{eq:J1series}
J_1(t;c_{\alpha},\delta) = \Bigg(\exp{\left(L_{\chi_{n}}\right)}
\ldots\exp{\left(L_{\chi_{2}}\right)}\exp{\left(L_{\chi_{1}}\right)} 
\left(J_R^{(n)}\right)\Bigg)\Bigg|_{J_R^{(n)}=J_0,\phi_R^{(n)}=
\phi_0,\phi_F^{(n)}=\omega_F t}~~,
\end{equation}
\begin{equation}\label{eq:J2series}
J_2(t;c_{\alpha},\delta) = \Bigg(\exp{\left(L_{\chi_{n}}\right)}
\ldots\exp{\left(L_{\chi_{2}}\right)}\exp{\left(L_{\chi_{1}}\right)} 
\left(J_F - \frac{k}{\ell}J_R^{(n)}\right)\Bigg)\Bigg|_{J_R^{(n)}
=J_0,\phi_R^{(n)}=\phi_0,\phi_F^{(n)}=\omega_F t}~~,
\end{equation}
\begin{equation}\label{eq:theta1series}
\phi_1(t;c_{\alpha},\delta) = \Bigg(\exp{\left(L_{\chi_{n}}\right)}
\ldots\exp{\left(L_{\chi_{2}}\right)}\exp{\left(L_{\chi_{1}}\right)} 
\left(\phi_R^{(n)}+\frac{k}{\ell} 
\phi_F^{(n)}\right)\Bigg)\Bigg|_{J_R^{(n)}=J_0,\phi_R^{(n)}
=\phi_0,\phi_F^{(n)}=\omega_F t}~~,
\end{equation}
\begin{equation}\label{eq:theta2series}
\phi_2(t;c_{\alpha},\delta) = \Bigg(\exp{\left(L_{\chi_{n}}\right)}
\ldots\exp{\left(L_{\chi_{2}}\right)}\exp{\left(L_{\chi_{1}}\right)} 
\left(\phi_F^{(n)}\right)\Bigg)\Bigg|_{J_R^{(n)}=J_0,\phi_R^{(n)}
=\phi_0,\phi_F^{(n)}=\omega_F t}~~.
\end{equation}
The equations \equ{eq:J1series}-\equ{eq:theta2series} provide the 
variation in time of the periodic orbit.

Moreover, the above equations give a generalised expression for the 
position of the periodic orbit with respect to the system parameters. 
Therefore, they could be used to compute the characteristic curves 
for the families of periodic solutions in the parameter space.

\subsection{Bifurcation thresholds}\label{sec:bif}
By varying the energy level or some control parameters, it can happen 
that the equilibrium solutions \eqref{equipoint} undergo a transition 
from stability to instability, or vice-versa. For topological reasons 
this phenomenon implies the appearance or disappearance of additional 
critical points with associated {\em bifurcations} of new families of 
periodic orbits. Usually, an analysis of the Hessian determinant in 
terms of internal and control parameters is straightforward and then 
it is possible to get explicit \sl bifurcation curves \rm in a relevant 
parameter space. The computation of the normal form allows one to refine 
the results and to obtain bifurcation values as close as possible to the 
curves computed by a numerical approach.

\subsection{Error estimates and optimal order}\label{sec:so}

The precision of the normal form is measured by the size of the remainder
function. Thus, it makes sense to measure the size of $R^{(n)}$ at each
order $n$ of the normalization procedure. This section is devoted to provide
formal norm definitions allowing to estimate the size 
of the remainder function.

With reference to \equ{rem}, let $R^{(n)}:\real^2\times \torus^2\rightarrow 
\real$ be the remainder function, that we write in the form
\beqa{remform}
R^{(n)}(J_1,J_2,\phi_2,\phi_2;c_\alpha,\lambda,\delta)&=&
\lambda^{n+1} H^{(n)}_{n+1} + \lambda^{n+2} H^{(n)}_{n+2}+\mathcal{O}_{n+3}\nonumber\\
&=&\sum_{s=1}^\infty \lambda^{n+s} \sum_{s_1,s_2,k_1,k_2}
a^{(n,s)}_{s_1 s_2 k_1 k_2} (c_{\alpha},\delta) J_1^{s_1\over 2}
J_2^{s_2\over 2} {\rm e}^{i (k_1 \phi_1 + k_2 \phi_2)}\ ,\nonumber\\
\eeqa
where $s_1,s_2 \in \mathbb N$ and $k_1,k_2 \in \integer$, with lower and
upper bounds depending on the order of the book-keeping $\lambda$.
The coefficients $a^{(n,s)}_{s_1 s_2 k_1 k_2}$ are computed via the 
recursive application of the Lie normalisation scheme 
(\eqref{lieham} and \eqref{eq:homological}).
For a sufficiently small parameter $\xi>0$, the Lie series procedure
guarantees that the series $R^{(n)}$ is convergent in a set
\beq{Gamma}
\Gamma\equiv \{(J_1,J_2,\phi_2,\phi_2)\in\real^2\times \torus^2:\
|J_i|<\xi\ ,\ \phi_i\in\torus\ ,\ i=1,2\}\ .
\eeq
The parameter $\xi$ gives a measure of the size of the 
domain in the actions around the equilibrium position, where the 
normal form method is applicable, i.e, the associated Lie
transformation converges (see Remark~\ref{remarkremainder} below). On the other hand,
when computing the normal form explicitly, possibly by means of an algebraic
manipulator, we need to truncate the series expansions appearing in \equ{remform}.
To this end, let $N$ be the order of the truncation, and let $R^{(n,N)}$
be the truncated remainder function defined as
\begin{equation}
R^{(n,N)} = \sum_{s=1}^N \lambda^{n+s} \sum_{s_1,s_2,k_1,k_2} a^{(n,s)}_{s_1 s_2 k_1 k_2}
(c_{\alpha},\delta) J_1^{s_1/2} J_2^{s_2/2} {\rm e}^{i (k_1 \phi_1 + k_2 \phi_2)}\ .
\end{equation}
Given that the function $R^{(n,N)}$ is still defined in the set $\Gamma$ as in 
\equ{Gamma}, we introduce the following \sl majorant \rm norm, which depends 
on the control parameters $c_{\alpha}$ as well as on the detuning $\delta$:
\begin{equation}\label{eq:remnormtrunc}
\|R^{(n,N)}\|_{(c_{\alpha},\delta,\xi)} = \sum_{s=1}^{N} \sum_{s_1,s_2,s_3,k_1,k_2}
\mid a^{(n,s)}_{s_1 s_2 k_1 k_2} (c_{\alpha},\delta) \mid \xi^{\frac{s_3}{2}}\ .
\end{equation}
Based on \equ{eq:remnormtrunc}, concrete analytical estimates of the size of 
the remainders $||R^{(n,N)}||$, at every order $n$, as well as the
optimal order, where $||R^{(n,N)}||$ becomes minimum, can be provided 
(see Section~\ref{sec:asym}).

\begin{remark}
$i)$ The sequence $\|R^{(n,N)}\|_{(c_{\alpha},\delta,\xi)}$, for fixed 
values of $n$, $c_{\alpha}$, $\delta$, $\xi$ and for $N=1,2,\ldots$ is 
convergent provided that $c_{\alpha}$, $\delta$, $\xi$ are sufficiently 
small (see, e.g., \cite{gior}). Its limit as $N\rightarrow \infty$ is
hereafter denoted $\|R^{(n,\infty)}\|$.

$ii)$ The sequence $\|R^{(n,\infty)}\|_{(c_{\alpha},\delta,\xi)}$ is 
{\sl asymptotic}. Indeed, a typical behavior is that for a normalization 
order $n$ small enough, the size $\|R^{(n,\infty)}\|_{(c_{\alpha},\delta,\xi)}$
decreases as $n$ increases. However, beyond a certain order which we refer to
as the {\sl optimal order}, say $n_{opt}$, the quantity 
$\|R^{(n,\infty)}\|_{(c_{\alpha},\delta,\xi)}$ starts to increase
with $n$. This shows that the minimum size of the remainder- corresponding 
to the best normal form approximation - occurs at the normalization order 
$n=n_{opt}$.

$iii)$ According to Nekhoroshev theory (see \cite{Nekh}, see also 
\cite{CHR}) the optimal order $n_{opt}$ decreases as the small parameters 
(e.g., $c_{\alpha}$, $\delta$ or $\xi$) increase.
\label{remarkremainder}
\end{remark}

\section{Application to the Secondary resonances of the synchronous 
resonance in the spin-orbit problem}\label{sec:model}

The general method described in the previous section is now applied 
to the particular cases of the secondary resonances of  the 1:1 primary 
resonance in the spin-orbit problem. More specifically, we study 
the three lowest order secondary resonances:  1:1,  2:1 and  
3:1. For each case we  construct a high-order normal form and 
provide a series of analytical computations. First, we  
compare the analytical Poincar\'e surfaces of section with the 
numerical ones, and confirm that our integrable approximation successfully 
captures the topological transitions accompanying the bifurcations of 
periodic orbits for each particular secondary resonance. 
Moreover, we compute the characteristic curves of the families 
of periodic orbits involved in the secondary resonances and compare 
them with those computed numerically by means of a Newton-Raphson method. 
Finally, the bifurcation curves for each resonance are determined 
analytically in the parameter space $(e,\delta)$.

\subsection{The 1:1 secondary resonance}
\label{subsec:p11s11}

The 1:1 secondary resonance becomes important for asphericities close to 
$\ep =1 $. This corresponds, e.g. to nearly prolate bodies with 
axial ratios $\approx 1.5$. Historically, the case of the satellite 
of Saturn Hyperion, which represents the first example of observationally
detected \emph{chaotic} spin rotation in the Solar system \cite{harbison}, belong to this class. Being, instead, interested in finding the various `modes' and parameters
for which ordered motion would be possible in the 1:1 secondary resonance,  
we follow the normalization procedure 
described in Sec.~\ref{sec:norma} and we consider
$$
k = \ell = 1.
$$
We apply the normalization scheme of Section~\ref{sec:gp} to 
the Hamiltonian \eqref{eq:hambookkept}, using a computer-algebraic 
program up to the normalization order $n=11$. The first few terms of
the normalized Hamiltonian read
$$
H = Z_0 + \lambda Z_1 + \lambda^2 Z_2 + \mathcal{O}_3
$$
with
\beqano
Z_0 &=& J + J_{\phi} \nonumber\\
Z_1 &=&  \delta J -   e \sqrt{2 J}  \cos( u - \phi)\ \\
Z_2 &=&  - 2 e^2 J - \frac{1}{4} J^2 - e^2 J \cos( u - \phi)+ 
\sqrt{2} e J^{3/2} \cos( u - \phi) - \frac{3}{2} e \sqrt{2J} 
\delta \cos( u - \phi).
\eeqano

Recall that, according to Remark~\ref{remarktransr}, $J$ and $u$ above denote the near identity transformation of the action angle variables defined in Eq.~\ref{AARV} after $n = 11$ normalization steps. 

Next, we introduce another set of canonical variables for the resonant Hamiltonian:
$$
\phi\rightarrow\phi_F\ ,\qquad u \rightarrow \phi_{R}+ \phi_{F}\ ,
\qquad J \rightarrow J_R\ ,\qquad J_{\phi} \rightarrow J_F - J_R\ .
$$
The transformed Hamiltonian becomes:
\beqano
H &= J_F + \lambda \delta J_R -2 \lambda^2  e^2 J^2_R - \frac{1}{4} \lambda^2  J_R^2 - \lambda e  \sqrt{2 J}  \cos( \phi_R) + \sqrt{2} \lambda^2 e J_R^{3/2} \cos( \phi_R) \\ 
&- \frac{3}{2} \lambda^2 e \sqrt{2J_R} \delta \cos(\phi_R)-  \lambda^2 e^2 J_R 
\cos( \phi_R) + \mathcal{O}_3.
\eeqano
We can further simplify the resonant Hamiltonian by applying a 
canonical transformation to Poincar\'e variables
\begin{equation*}
X = \sqrt{2 J_R} \sin{\phi_R}\ ,\quad Y = \sqrt{2 J_R} \cos{\phi_R}\ .
\end{equation*}
Since $J_F$ plays now the role of the dummy action $J_{\phi}$, without loss
of generality we can set $J_F=0$. Dropping the formal dependence on the book-keeping parameter $\lambda$ (see Remark~\ref{remarklambda}) the Hamiltonian in polynomial form reads:
\beqa{eq:HXYP11S11}
H &= \frac{1}{2} \delta (X^2 + Y^2) - \frac{1}{16} (X^4 + Y^4) - 
\frac{1}{2} e^2 X^2 - e Y + \frac{1}{2} e X^2 Y - \frac{3}{2} e^2 Y^2 \\
&-\frac{1}{8} X^2 Y^2 + \frac{1}{2} e Y^3 - \frac{3}{2} e \delta  Y  + \mathcal{O}_3 \nonumber 
\eeqa
The complete form of the function in Eq.~\eqref{eq:HXYP11S11} up to 
order $\mathcal{O}_6$ is given in the Appendix A. Since
the model \eqref{eq:HXYP11S11} is integrable, this allows to find
explicit analytical formulas approximating the time evolution of the
spin state in the domain of regular motion.

\subsubsection{Poincar\'e Surfaces of Section}

\begin{figure}
\centering
\includegraphics[width=0.8\textwidth]{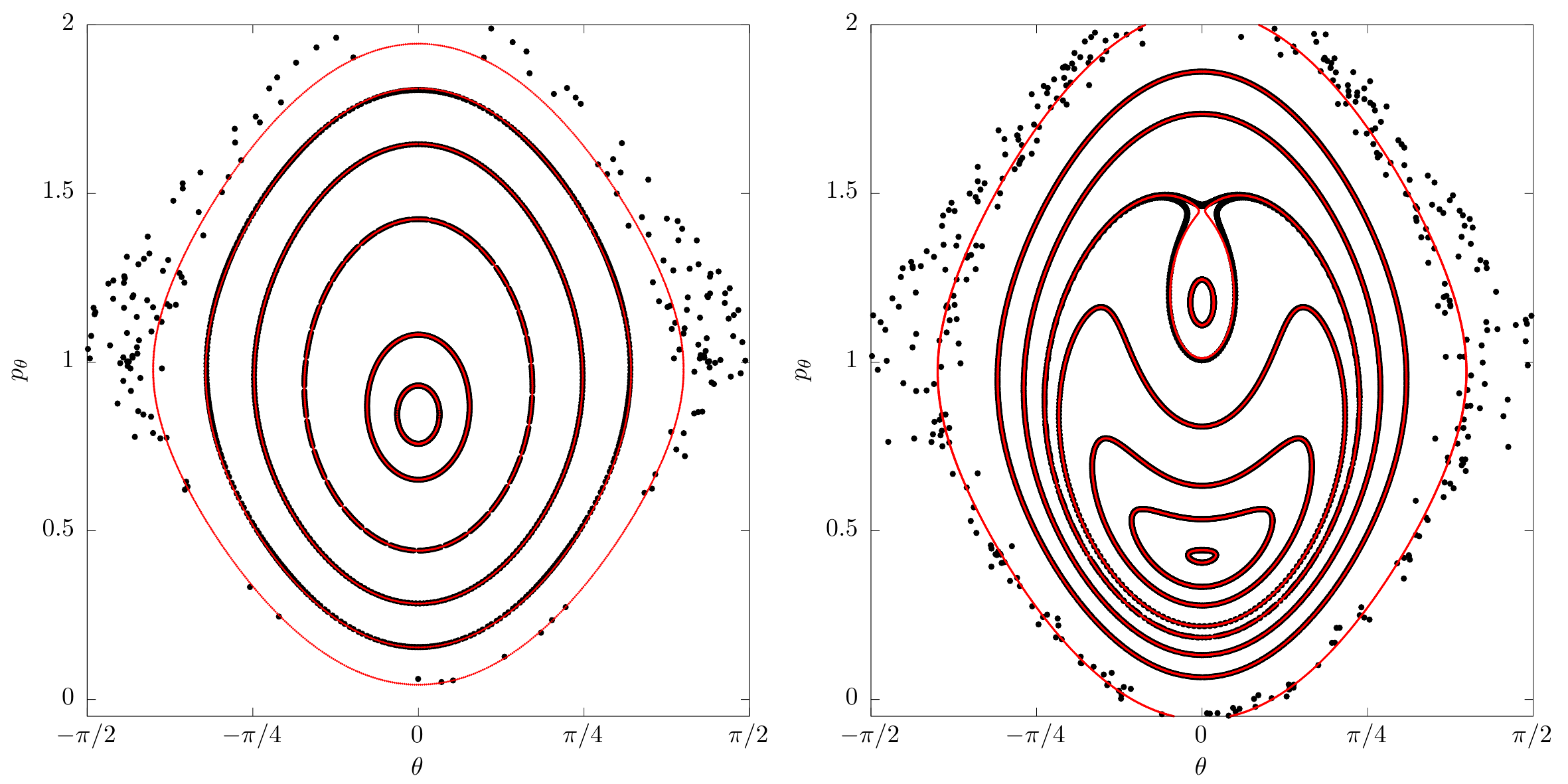}
\caption{Poincar\'e surfaces of sections for different values of 
the control parameters $(\ep,e)$: left panel $(0.93,0.01)$, right 
panel $(1.07,0.01)$. The sections produced from the level curves of the 
resonant Hamiltonian normal form truncated at the normalization order 11 
(red curves) are superposed to those produced from the numerical 
integration of the equations of motion (black points).}
\label{fig:PSSP11S11}
\end{figure}

In Fig.~\ref{fig:PSSP11S11} we superpose the analytically found 
invariant curves (red curves) to the numerical phase portrait 
(black dots) computed as a stroboscopic surface of section for the 
1:1 secondary resonance. The red curves correspond to level curves 
of constant energy of the Hamiltonian $\eqref{eq:HXYP11S11}$ 
back-transformed to the original variables. One sees that,
for values of the asphericity $\ep>1$ there can exist 
more than one synchronous state. At the point $(\ep=1,e=0)$ a tangent
bifurcation occurs and we have the appearance of a new pair of periodic 
solutions, one stable and one unstable. The topological changes in the phase space 
around this critical value are depicted in Fig.~\ref{fig:PSSP11S11}. 
For values of the parameters $(\ep=0.93,e=0.01)$, in the surface of 
section we observe a typical pendulum-like structure in the synchronous 
resonant domain. However, as we increase 
the asphericity to a value $\ep>1$, the phase portrait 
shows that two stable synchronous solutions co-exist. 
The lower stable solution is called 
the \emph{$\alpha$-mode} while the upper one is the \emph{$\beta$-mode} \cite{MS}. 
Both the $\alpha$ and $\beta$ mode are surrounded by the separatrix 
stemming from the third unstable solution.

We mention here that such a phase portrait corresponds to 
the so-called Second Fundamental Model of a resonance \cite{SFM}. In fact,
the resonant normalized Hamiltonian Eq.~\eqref{eq:HXYP11S11}
has indeed the form of the Second Fundamental Model, thus, allowing to describe
in a straightforward way the bifurcation to the $\beta$-mode. 

\subsubsection{Characteristic curves and bifurcation diagram}

\begin{figure}
\centering
\includegraphics[width=0.48\textwidth]{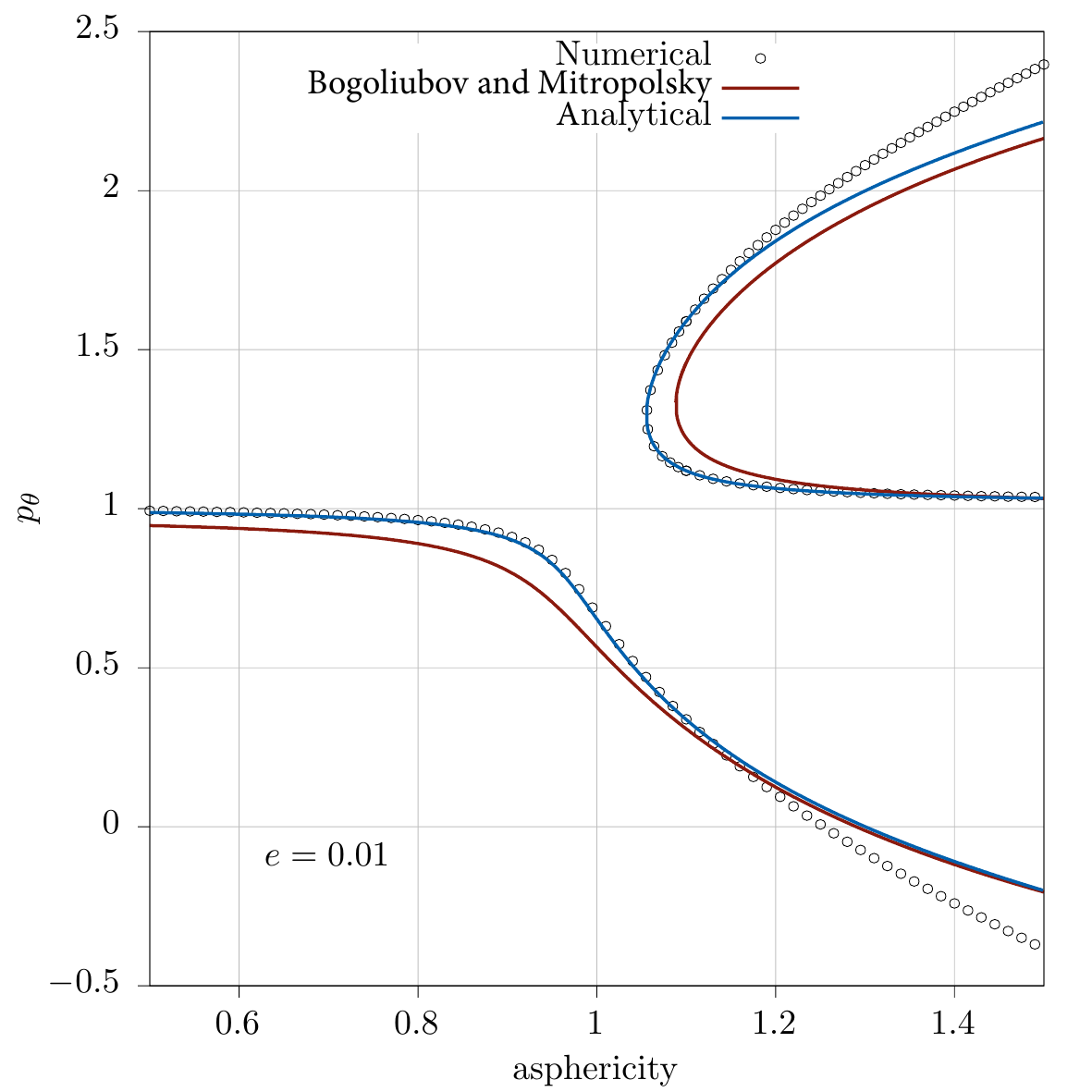}
\includegraphics[width=0.48\textwidth]{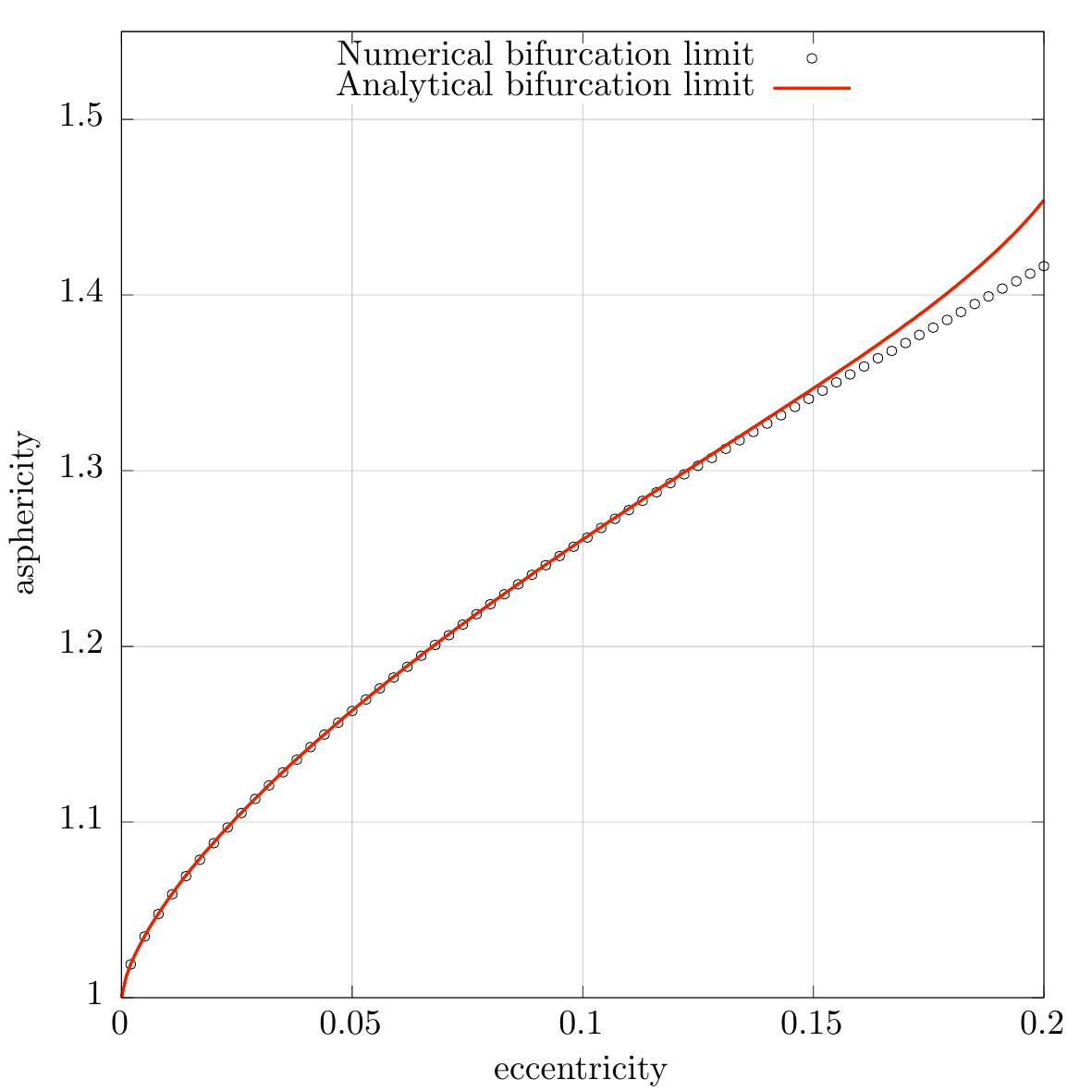}
\caption{On the left: the analytical estimates for the position of the periodic 
solution of the $\alpha$ and $\beta$-mode derived by our 11th order 
normal form construction (blue curve) and by the nonlinear
method of Bogoliubov and Mitropolsky \cite{bog1961,Wisdom1984} (red curve). 
The analytical results are also compared with those 
derived by the numerical method (black curve). On the right: 
the analytical estimates for the bifurcation curves derived 
by our 11th order normal form construction (red curve) and by the 
numerical method (black curve).}
\label{fig:P11S11char}
\end{figure}

The normal form construction allows to compute 
the characteristic curves (coordinates of the fixed point of the $\alpha$ 
and $\beta$ mode as one parameter is varied) for a given 
value of the eccentricity, and varying $\delta$, or vice versa. 
Fig.~\ref{fig:P11S11char} shows an example, for fixed $e=0.01$.
The periodic solutions are given as 
equilibrium points of the equations of motion derived by the resonant 
normalised Hamiltonian. For fixed $\ep,e$ one can solve the algebraic 
equation to find the equilibria and then back-transform them to the 
original variables. Fig.~\ref{fig:P11S11char} shows the excellent agreement
between the numerical\footnote{The numerical method uses the equations of 
motion derived by Eq.~\eqref{eq:fullham} and locates the synchronous 
periodic orbits via a Newton-Raphson process over the stroboscopic map.} 
and analytical characteristic curves. Note that $\delta$ is valued in the 
interval $-0.2<\delta<0.2$, which is about 20\% of the asphericity value 
$\ep=1$, corresponding to the central value of the secondary resonance.

As a comparison, another analytical method to estimate the position 
of the periodic orbits was proposed in \cite{Wisdom1984} using the nonlinear
method of Bogoliubov and Mitropolsky \cite{bog1961}. 
They derived a formula for the position of the synchronous resonance
\begin{equation}
p_{\theta} = (1 + \psi) \frac{1+e^2}{(1-e^2)^{3/2}},
\label{eqc2:wisdom1984}
\end{equation}
where $\psi$ is determined from the equation
$$
\psi - \ep^2 J_1(2 \psi) + 4 e =0,
$$
where $J_n$ are the usual Bessel functions. For values 
of $\ep>1$, Eq.~\eqref{eqc2:wisdom1984} has 1 or 3 solutions depending on 
the eccentricity, allowing us to compute the positions of the $\alpha$ and 
$\beta$-modes. The solution is also shown in Fig.~\ref{fig:P11S11char}. 
The purely numerical method has the Newton-Raphson accuracy $10^{-13}$ 
(black curves). The red and blue curves are computed, respectively,
using the analytical formula provided by \cite{Wisdom1984} and  using 
our resonant normal form up to the normalization order $n=11$. 
Our analytical estimates give the best agreement with the 
numerical computations. We note that, the analytical estimate 
for the position of the $\alpha$-mode and 
$\beta$-mode is satisfactory not only very close to the value of 
the asphericity $\ep = 1$, but also in a significant interval $[0.8,1.2]$  
around it. In fact, the $\alpha$-mode is very well represented from values 
of $\ep$ from about 0.5 up to about 1.2 where the normal form solution 
starts diverging. However, it is interesting the fact that although 
diverging from the numerical solution, the normal form estimate now 
converges to the other analytical estimate from Wisdom's formula.
Since all these normal form constructions are supposed to work well
in local domains (in the actions or the parameters, see section~\ref{sec:gp}), 
we suspect that the similarity observed in the divergence of the two analytical 
predictions is related to the overall expected failure of the averaging process 
(performed either with the nonlinear
method of Bogoliubov and Mitropolsky \cite{bog1961,Wisdom1984} or our proposed 
normal form method) in a range of parameters outside this domain.

Finally, the right panel of Fig.~\ref{fig:P11S11char} shows
the computation of the complete \emph{bifurcation diagram} of 
the tangent bifurcation. As already mentioned, topological 
transitions in the phase portrait are associated with the appearance of 
a pair of new periodic solutions that appears along the 
$\theta = 0$ axis. The periodic solutions of the system correspond 
to fixed points of the normal form. Moreover, in Poincar\'e variables 
the Hamiltonian has a polynomial form and for $\theta=0$ we can set 
$X=0$. Then it suffices to study the number of real roots of the 
polynomial $H(Y)$: the points in the $(e,\ep)$-plane where we 
pass from 1 to 3 real roots give us the analytical locus of the 
bifurcation curve. In the same manner, one can do the same computation 
numerically by finding the set of points in the $(e,\ep)$-plane 
where we pass from one periodic solution to three. The results show 
that the analytical predictions fit well with the numerical ones up to 
$e\approx 0.15$, $\delta\approx 1.3$. Again here the limits are connected
with the domain of applicability of the normal form approach,
and they are further commented in section~\ref{sec:asym}, where
a detailed analysis of the error of the method is made.  

\subsection{The 2:1 secondary resonance}

The normal form construction of the 2:1 secondary resonance of the 
synchronous primary resonance is presented 
in detail in \cite{MNRASpaper}. We summarize here some basic results,
and proceed in a detailed error analysis for this resonance in 
Section~\ref{sec:asym}. We have $$k=1 , \quad \ell = 2 ,$$
and the normalized Hamiltonian reads 
$$
H = Z_0 + \lambda Z_1 + \lambda^2 Z_2 + \mathcal{O}_3
$$
with $Z_0 = \frac{1}{2} J + J_{\phi}$,  
$Z_1 =  \delta J - \frac{3}{8}  e J \cos(2 u - \phi)$ and
$Z_2 =  \frac{89}{128} e^2 J- \frac{1}{4} J^2 - \frac{3}{4} 
e \delta J \cos(2 u - \phi)$. In the resonant variables
$$
\phi\rightarrow\phi_F\ ,\qquad u \rightarrow \phi_{R}
+\frac{1}{2} \phi_{F}\ ,\qquad J \rightarrow J_R\ ,
\qquad J_{\phi} \rightarrow J_F - \frac{1}{2} J_R\ ,
$$
the normal form becomes:
$$
H = J_F + \lambda \delta J_R - \frac{89}{128}  \lambda^2 e^2 J_R - 
\frac{1}{4} \lambda^2 J_R^2 -   e J_R \left(  \frac{3}{8} \lambda  + 
\frac{3}{4} \lambda^2 \delta \right) \cos(2 \phi_R) + \mathcal{O}_3.
$$
In the Poincar\'e variables $X = \sqrt{2 J_R} \sin{\phi_R}\ ,\quad Y = \sqrt{2 J_R} \cos{\phi_R}$, and setting as before $J_F=0$, one gets the 
Hamiltonian in a polynomial form:
\begin{eqnarray}\label{HXY11} 
H = \frac{3}{16} e X^2 - \frac{89}{256} e^2 X^2 -
 \frac{1}{16} X^4 -  \frac{3}{16} e Y^2 - \frac{89}{256} e^2 Y^2 
- \frac{1}{8} X^2 Y^2 -
 \frac{1}{16} Y^4 + \nonumber \\ 
 + \frac{1}{2}  X^2 \delta +
 \frac{3}{8} e X^2 \delta + \frac{1}{2} Y^2 \delta - \frac{3}{8} 
e Y^2 \delta + \mathcal{O}_3.
\end{eqnarray}
The complete form of the function in Eq. (\ref{HXY11}) up to order 
${O}_6$ is given in the Appendix (see also \cite{MNRASpaper}).
Similarly to the case of the 1:1 secondary resonance, these 
explicit formulas can be used to derive analytical approximations
for the time evolution of the spin state in the domain of 
applicability of the normal form.

\begin{figure}
\includegraphics[width=0.95\textwidth]{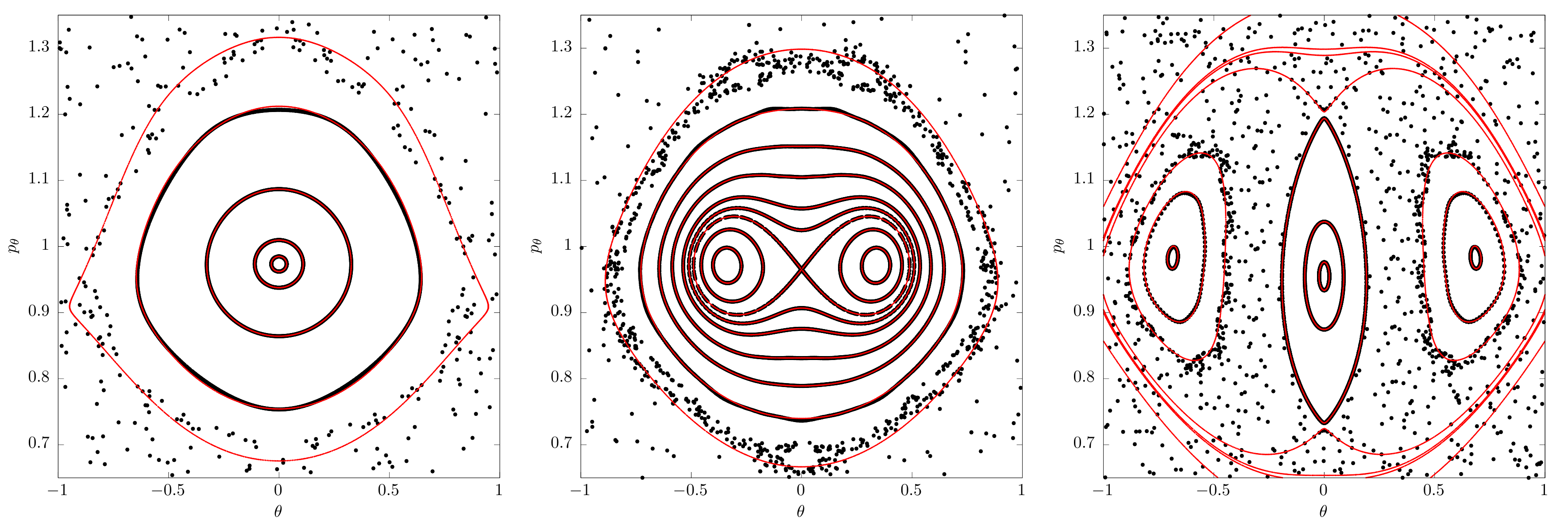}
\caption{Poincar\'e surfaces of sections for $e=0.05$ and for 
different values of the asphericity: left panel $(\ep = 0.45)$, 
central panel $(\ep = 0.5)$ and right panel $(\ep = 0.55)$. The 
sections produced from the level curves of the resonant Hamiltonian 
normal form truncated at the normalization order 11 (red curves) are 
superposed to those produced from the numerical integration of the 
equations of motion (black points).}
\label{fig:PSSP11S21}
\end{figure}
\begin{figure}
\includegraphics[height=0.3\textheight]{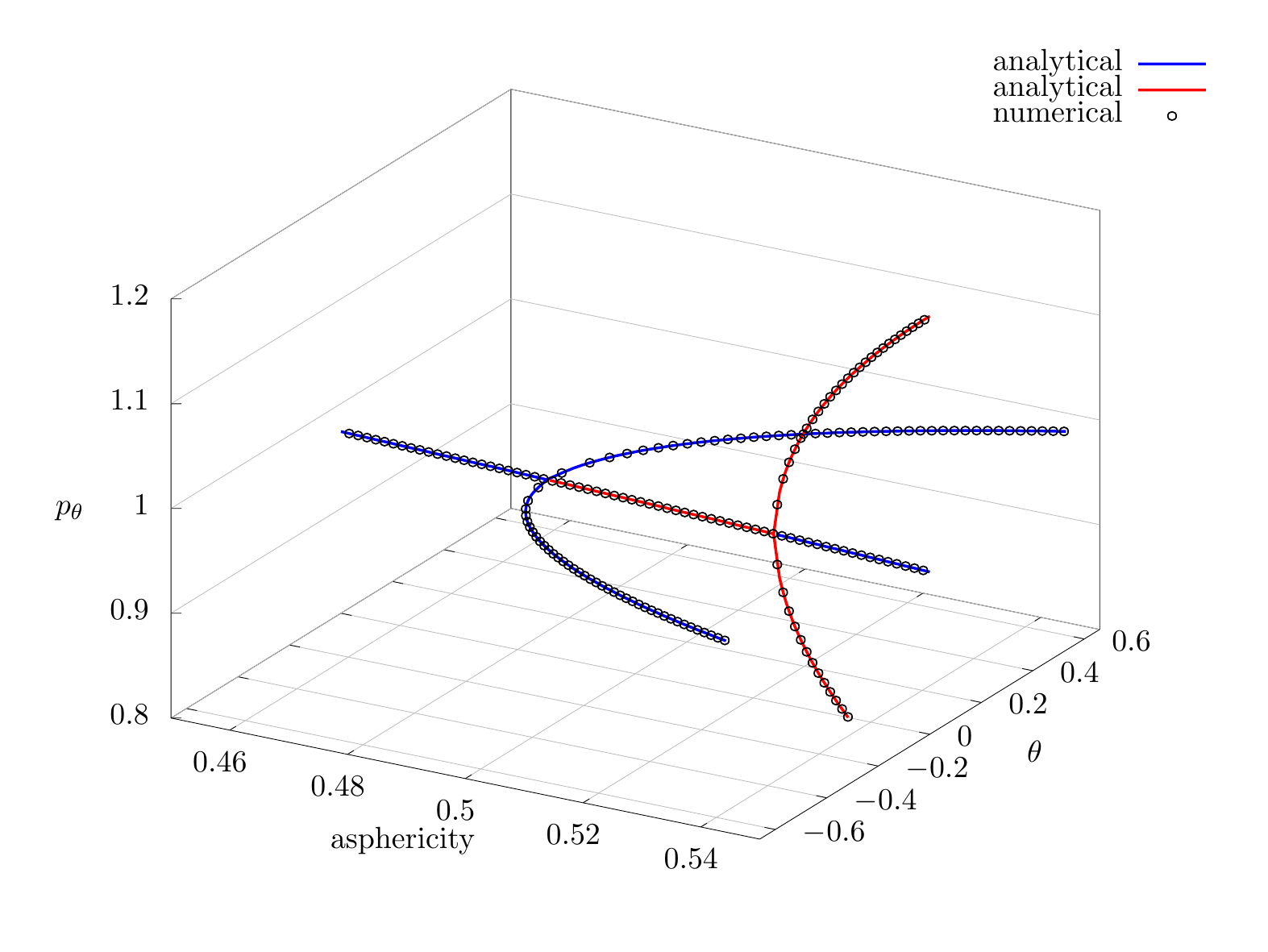}
\includegraphics[height=0.3\textheight]{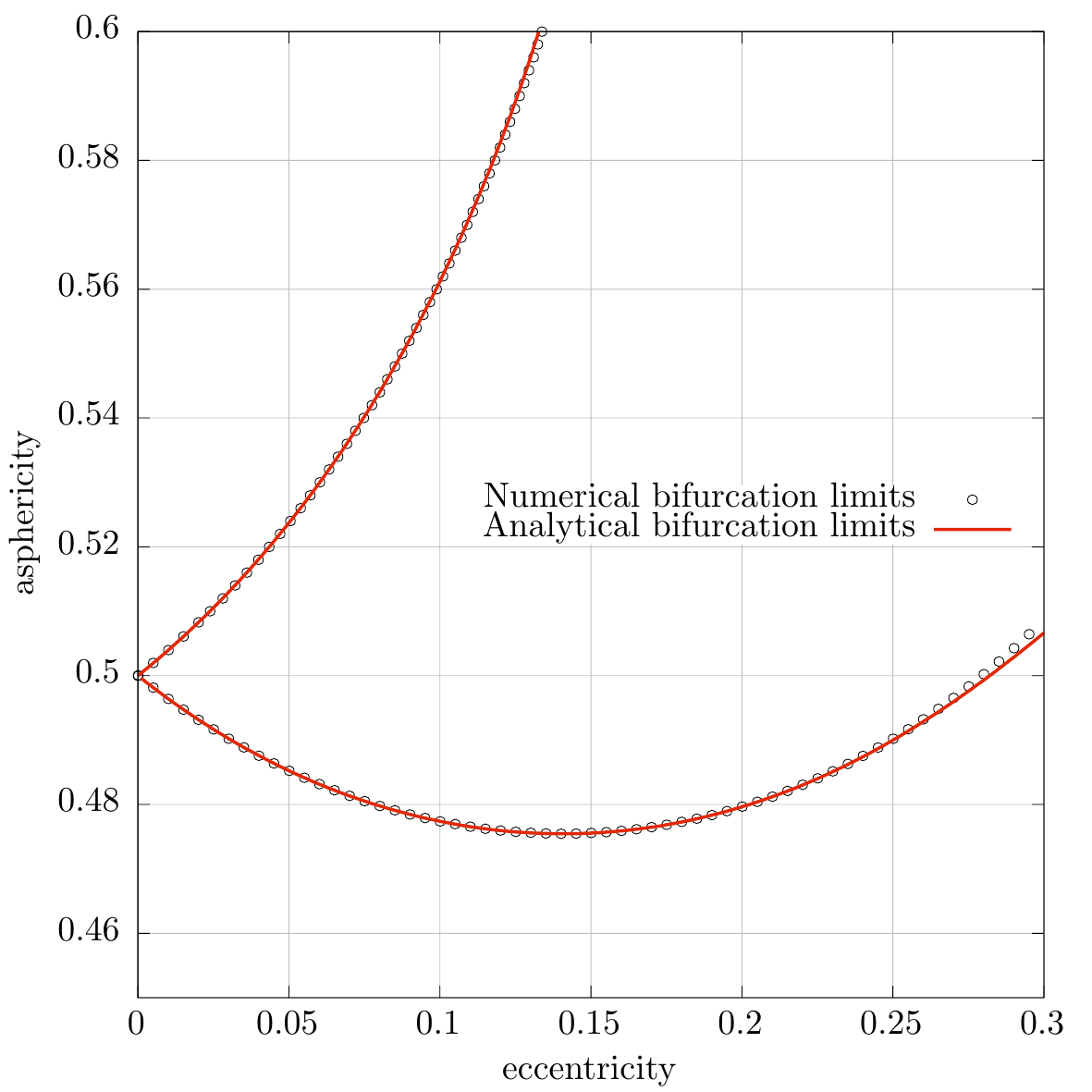}
\caption{Left panel: the analytical estimates for the position of the periodic 
solutions involved in the 2:1 secondary resonance. The colors for the 
analytical solutions denote the stable (blue) and unstable (red) members of each family.
The numerical results are superposed with black circles. Right panel:
the bifurcation diagram for the 2:1 resonance.}
\label{fig:P11S21charbif}
\end{figure}

The transitions in the phase-space of the spin-orbit problem in the 
case of the 2:1 secondary resonance were studied in \cite{MNRASpaper}, while a further 
example is shown in Fig.~\ref{fig:PSSP11S21}.
Note that even as chaos increases fast as the 
eccentricity increases ($e=0.05$ in Fig.~\ref{fig:PSSP11S21}), the invariant curves found by the 
normal form capture precisely the dynamics in places where regular islands 
still exist. Hence, the normal form reproduces well the bifurcations of 
periodic orbits around the primary resonance. In particular, the system undergoes two critical transitions. 
First, the primary resonance becomes unstable and we have the appearance 
of a pure figure-8 structure (Fig.~\ref{fig:PSSP11S21} central panel).
A stable family of periodic orbits appears on either side of the central 
resonance for almost the same value of the action $p_\theta$. 
By further changing the control parameter, we have another topological 
transition. The central resonance becomes stable again, and two unstable 
periodic orbits appear for the same value of the angle $\theta$ 
(Fig.~\ref{fig:PSSP11S21} right panel). 

The characteristic curves, showing these transitions, are depicted in left panel of 
Fig.~\ref{fig:P11S21charbif}. We compute analytically the characteristic 
curves for the families of periodic orbits involved in the topology of 
the 2:1 secondary resonance. The stability of each periodic orbit
is also computed from the eigenvalues of the linearised matrix for 
each equilibrium solution. The two families of stable (blue) periodic 
orbits that appear on the first bifurcation and the two families 
of unstable (red) periodic orbits are presented, along with the 
central periodic orbit. Moreover, we can estimate the threshold 
of the two critical transitions in the topology, both analytically
and numerically. The results are presented in the right panel of 
Fig.~\ref{fig:P11S21charbif}. For more details on these computations we refer 
the reader to \cite{MNRASpaper}.

\subsection{The 3:1 secondary resonance}
%
In the case of the 3:1 secondary resonance we follow the normalization 
procedure described in Section~\ref{sec:gp} with 
$$
k=1 , \quad \ell = 3.
$$  
By applying the above normalization scheme on the Hamiltonian 
(\ref{eq:hambookkept}), the normalized Hamiltonian reads
$$
H = Z_0 + \lambda Z_1 + \lambda^2 Z_2 + \mathcal{O}_3
$$
with
\beqano
Z_0 &=& \frac{1}{3} J + J_{\phi} \nonumber\\
Z_1 &=& \delta J  \\
Z_2 &=& - \frac{4}{15} e^2 J - \frac{1}{4} J^2 - 
\sqrt{\frac{2}{27}} e J^{3/2} \cos(3 u - \phi) .
\eeqano
Introducing the resonant canonical variables:
$$
\phi\rightarrow\phi_F\ ,\qquad u \rightarrow \phi_{R}+\frac{1}{3} 
\phi_{F}\ ,\qquad J \rightarrow J_R\ ,\qquad J_{\phi} 
\rightarrow J_F - \frac{1}{3} J_R\ ,
$$
the transformed Hamiltonian becomes:
$$
H = J_F + \lambda \delta  J_R - \frac{1}{4} \lambda^2 J_R^2 -\frac{4}{15} \lambda^2 e^2 J_R 
- \sqrt{\frac{2}{27}} \lambda^2 e J_R^{3/2} \cos(3 \phi_R) + \mathcal{O}_3,
$$
or, in Poincar\'e variables $X = \sqrt{2 J_R} \sin{\phi_R}\ ,\quad Y = \sqrt{2 J_R} \cos{\phi_R}$ (with $J_F=0$):
\beqano\label{HXYP11S31} 
H =& \frac{1}{2} \delta (X^2+Y^2) - \frac{1}{16} (X^4+Y^4) - 
\frac{2}{15} e^2 (X^2 + Y^2) - \frac{1}{8} X^2 Y^2  \\
&- \frac{1}{6 \sqrt{3}} e Y^3 + \frac{1}{2\sqrt{3}} e X^2 Y  + \mathcal{O}_3.
\eeqano
The complete form of the function in Eq. \eqref{HXYP11S31} up to 
order $\mathcal{O}_6$ is given in the Appendix. 

\begin{figure}
\centering
\includegraphics[width=\columnwidth]{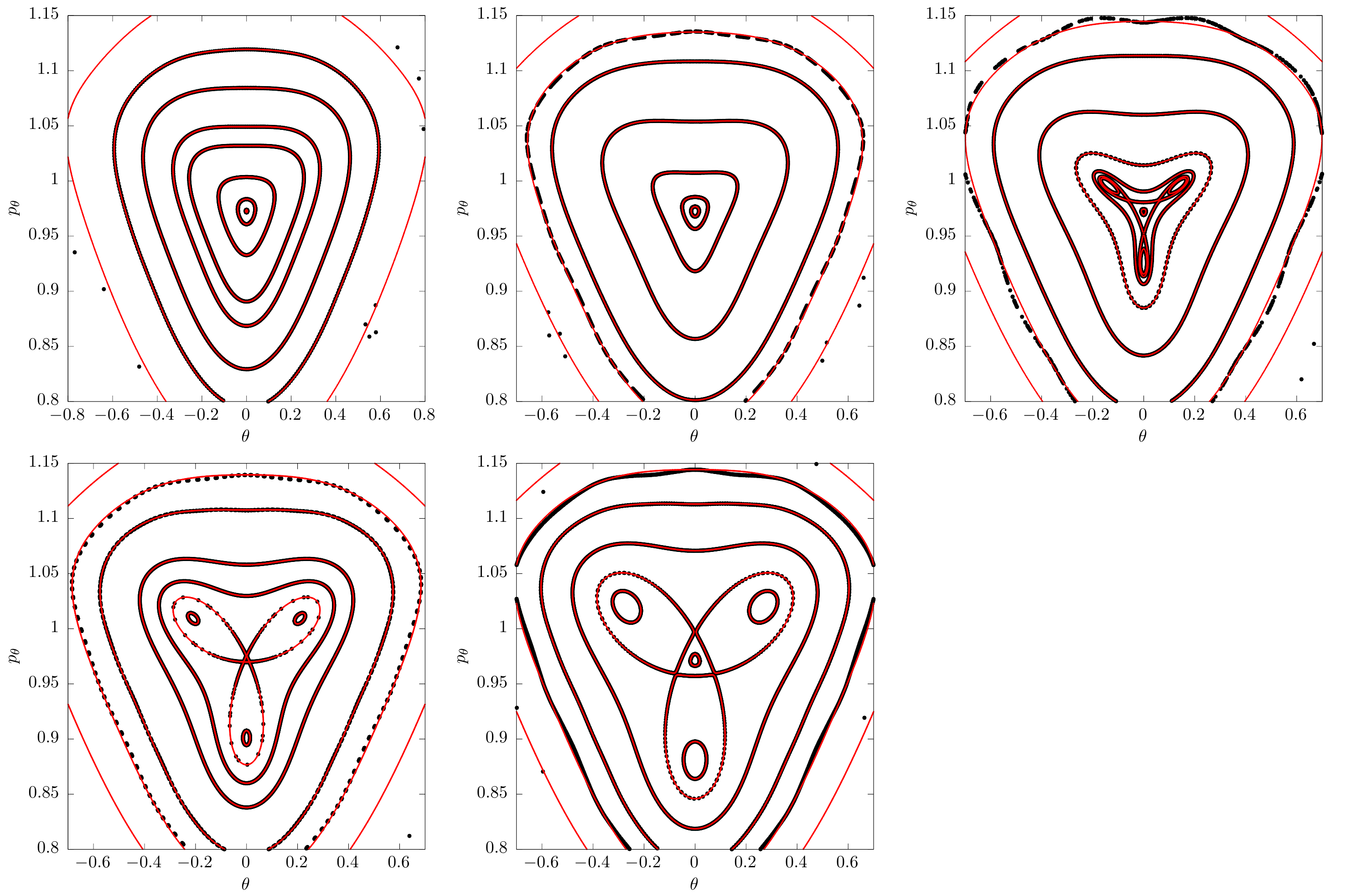}
\caption{Poincar\'e surfaces of sections for $e=0.1$ and for different 
values of the asphericity $\ep$: (from top-left to bottom-right) 
$\ep = 0.3333$, $\ep = 0.3350$, $\ep = 0.3370$,$\ep = 0.3381$ and 
$\ep = 0.34$. The sections produced from the level curves of the resonant 
Hamiltonian normal form truncated at the normalization order 
11 (red curves) are superposed to those produced from the numerical 
integration of the equations of motion (black points).}
\label{fig:PSSP11S31}
\end{figure}

The topology around the 1:1 primary resonance changes dramatically 
as we approach the critical value of the asphericity $\ep = 1/3$. 
In Fig.~\ref{fig:PSSP11S31} we present a series of Poincar\'e surfaces 
of section that try to capture all the possible transitions. These 
transitions take place as $\delta$ is varied by about $\delta=0.01$.
At first, the primary resonance yields the well-known center topology 
(Fig.~\ref{fig:PSSP11S31} top-left panel). As we approach the 
critical value of $\ep$ for the appearance of the secondary 
resonance the inner region of the resonance takes a triangle 
shape pointing downwards (Fig.~\ref{fig:PSSP11S31} top-centre 
panel). Then a chain of islands of period 
3 appears on the edges of this triangle (Fig.~\ref{fig:PSSP11S31} 
top-right panel). The central periodic orbit is still surrounded by 
the separatrix created by the period-3 unstable periodic orbit. 
The separatrix keeps shrinking until it actually coincides with the 
central orbit. At this point, the so-called squizing effect happens. 
Further increasing the $\ep$ value, the separatrix appear again, 
with the same triangular shape, but this times it looks upwards 
(Fig.~\ref{fig:PSSP11S31} bottom-left panel). Finally the 3:1 
secondary resonance moves away from the central one, which takes 
its regular shape again (Fig.~\ref{fig:PSSP11S31} bottom-right 
panel). This peculiar chain of bifurcations in the 3:1 resonance is well known (see Appendix 7 of \cite{arn1978}).

The left panel of Fig.~\ref{fig:P11S31bifchar} shows the 
analytically computed characteristic 
curves for the families of the periodic orbits involved in the 
topological transitions of the 3:1 secondary resonance. Both the 
new appearing stable and unstable families are of multiplicity 3.
For clarity, in the left panel Fig.~\ref{fig:P11S31bifchar} we 
present only the initial conditions with respect to the action value 
$p_\theta$ and $\theta=0$.

\begin{figure}
\centering
\includegraphics[width=0.45\textwidth]{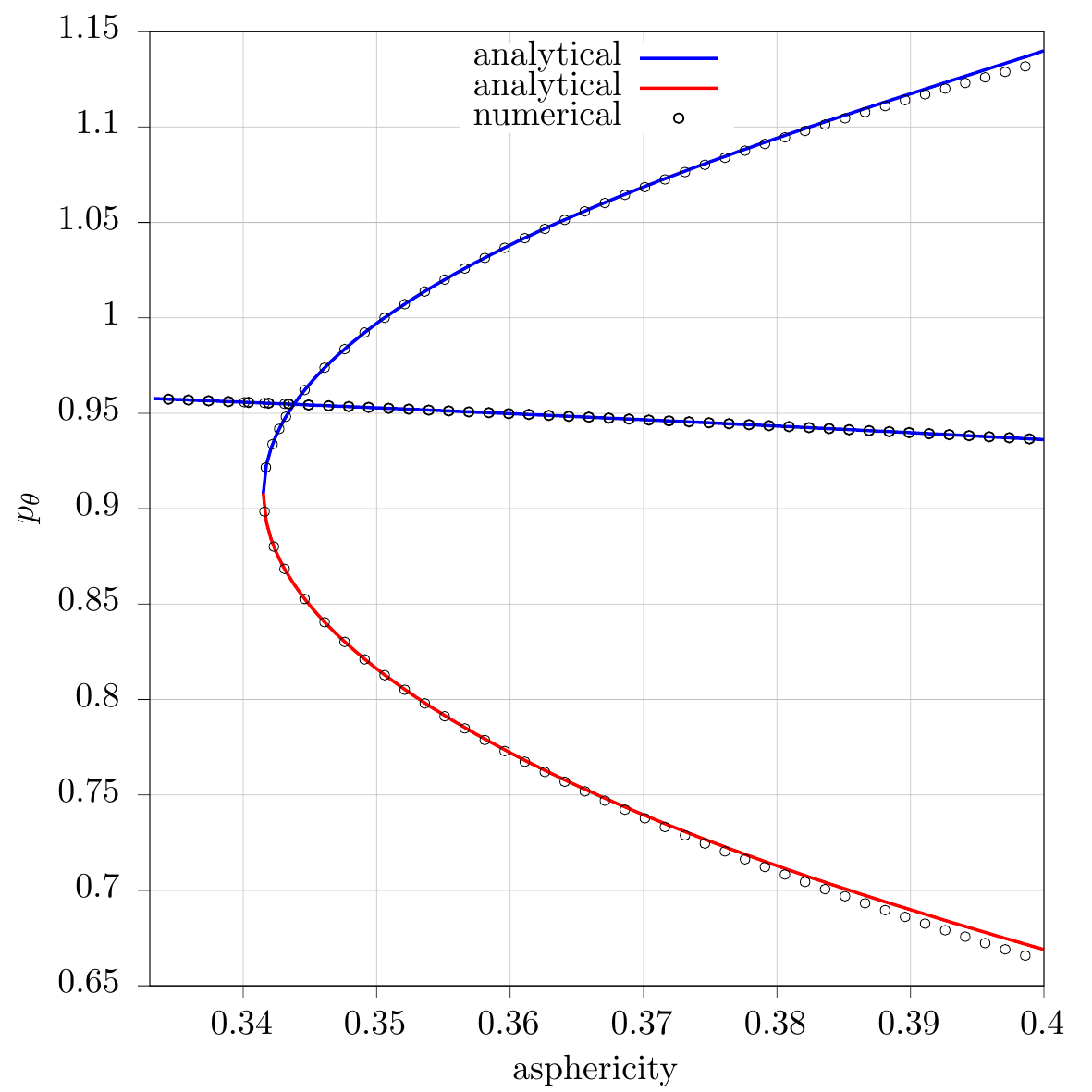}
\includegraphics[width=0.45\textwidth]{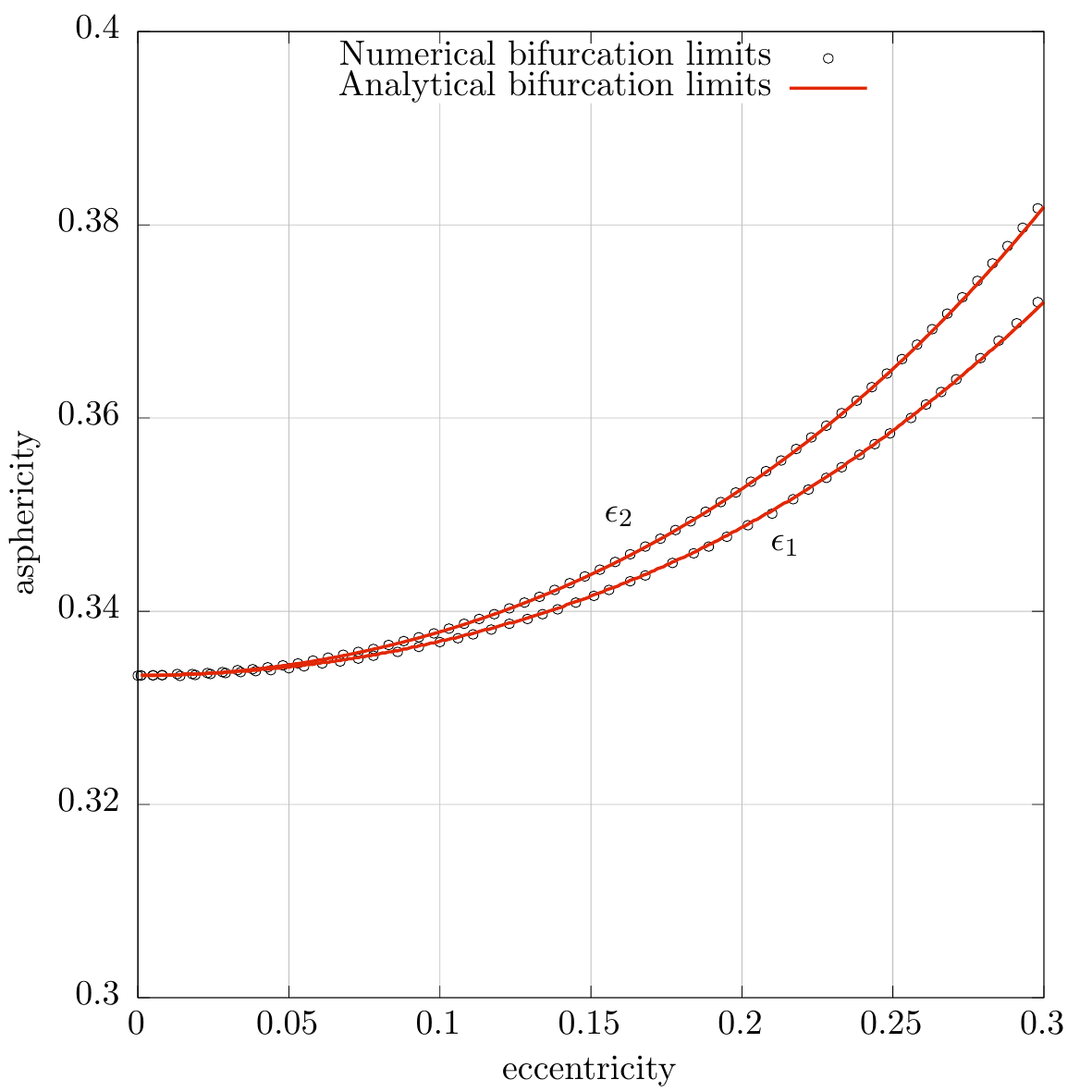}
\caption{On the left: the analytically computed characteristic curves for 
the periodic solutions associated with the 3:1 secondary 
resonance. On the right: bifurcation curves for the 3:1 secondary resonance of 
the 1:1 primary. The bifurcation limit $\ep_1$ corresponds to the 
values of the parameters where the chain of period 3 island chain 
appears. Bifurcation limit $\ep_2$ corresponds to the passage 
of the secondary resonance through the main one.}
\label{fig:P11S31bifchar}
\end{figure}

A determination of bifurcation curves in the parameter space 
$(\ep,e)$, where the period-3 chain of 
resonant islands around the primary resonance appears, can
be done as follows. Numerically this 
is detected by looking for period-3 solutions in the vicinity 
of the primary resonance. We fix one of the parameters and we smoothly 
change the other until we encounter the period-3 solution for 
the first time. Analytically the same work can be done, using 
the normal form for the secondary resonance. Now,
we count the number of roots of the multivariable 
polynomial in Poincar\'e variables. The results of this calculation
are presented in the right panel of Fig.~\ref{fig:P11S31bifchar}, 
where the above described bifurcation limit is denoted $\ep_1$. 

However, the topology is much more rich than a single 
bifurcation. As the value of the asphericity continues to increase,
we have the secondary resonance to pass through the primary resonance, 
which for an instance becomes unstable. This phenomenon can also be 
studied with our theory and the bifurcation curve is shown in the right 
panel of Fig.~\ref{fig:P11S31bifchar} as $\ep_2$. We estimate 
both analytically and numerically this limit by looking at the stability 
properties of the primary resonance.

\section{Series Asymptotic behavior and Error Analysis}\label{sec:asym}
\begin{figure}
\centering
\includegraphics[width=0.7\textwidth]{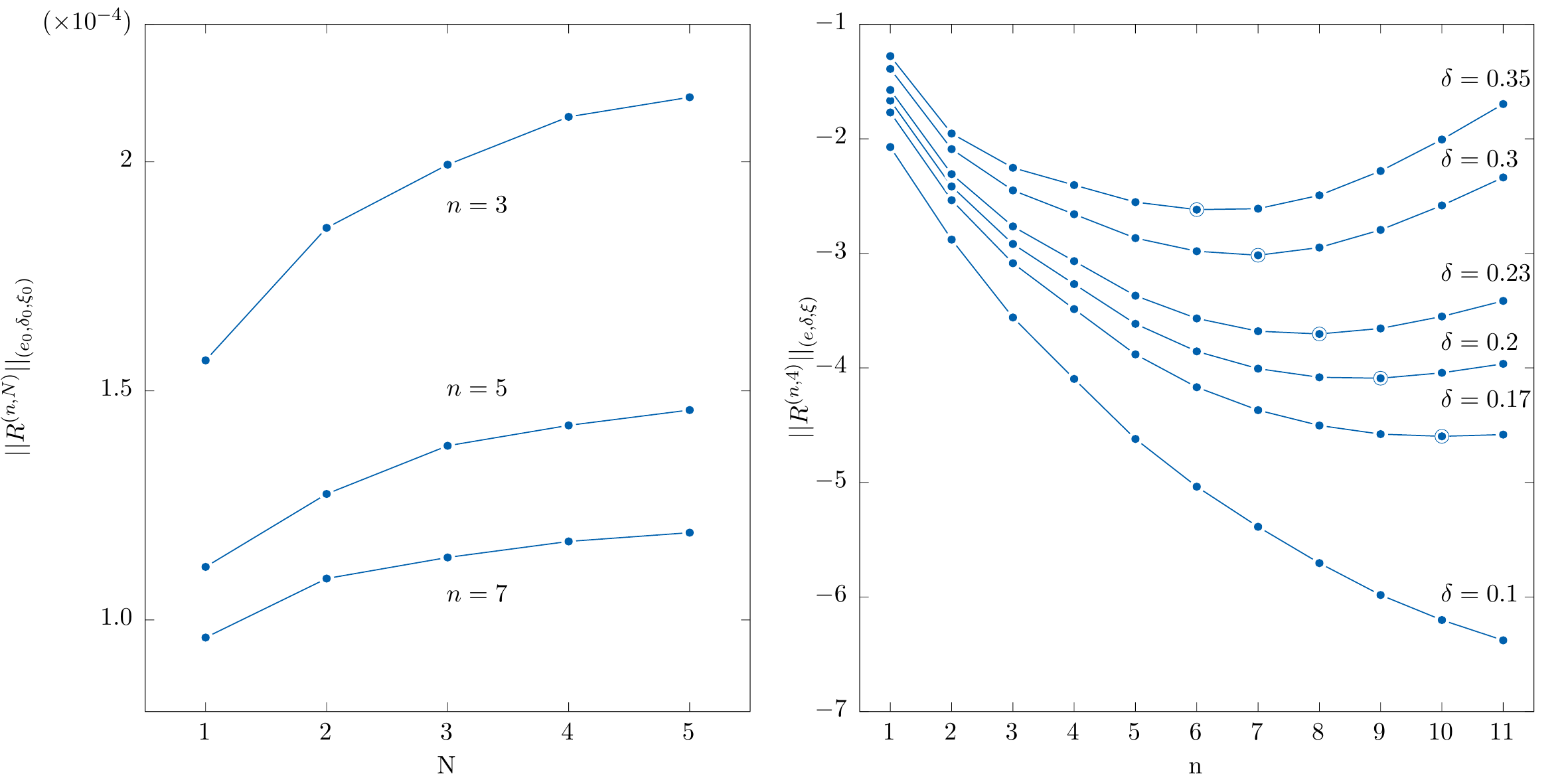}
\caption{In the left panel we observe the asymptotic behavior of the 
remainder function for different normalization orders $n$. The error 
saturates as a function of the number of terms $N$ in the truncated 
remainder function. In the right panel the optimal normalisation order,
denoted by a circularized point, is presented for different values 
of the control parameters.}
    \label{fig:ASYM}
\end{figure}

In this section we apply the error analysis estimates introduced in 
subsection (3.8), based on the asymptotic behavior of the remainder 
function associated with the normal forms computed in the previous 
sections. The basic quantity of interest is 
$\|R^{(n,N)}\|_{(c_{\alpha},\delta,\xi)}$, introduced in 
Eq.(\ref{eq:remnormtrunc}). Given particular parameter values $e,\delta$, 
the first step in the analysis is to check that the successive normalizations 
keep our transformed Hamiltonian convergent within the domain $|J_i|<\xi$, 
for a value of $\xi$ selected so as to contain all orbits which we are 
interested in. Figure {\ref{fig:ASYM}} (left panel) gives an example of 
such testing: The quantity $\|R^{(n,N)}\|_{(e,\delta,\xi)}\|$ is computed 
in the case of the $2:1$ secondary resonance, for $e=0.01$, $\delta = 0.1$ 
$\xi = 0.01$, and three different normalization orders, $n=3,5$ and $7$. 
In all three cases, the truncated remainder norm is computed when the 
truncation order extends to $N=n+q$, with $q=1,\ldots , 5$. One sees a 
rapid convergence of the remainder norm to a limiting value: actually, 
with a truncation even as low as $q=1$ one obtains a remainder value 
estimate which is, within a factor smaller than 2, close to the 
limiting value. We emphasize that this {\it convergence} test is crucial: 
contrary to a widespread belief, for the analytical approach to be valid, 
all performed normalizations must lead to convergent expressions as regards 
both the resulting canonical transformations and Hamiltonian normal form 
series. The celebrated `divergence' of the Birkhoff normal form 
refers to the divergence of the sequence
$$
\|R^{(n)}\|_{(e,\delta,\xi)}\equiv 
\lim_{N\rightarrow\infty}\|R^{(n,N)}\|_{(e,\delta,\xi)}\|,
$$
when the normalization order $n$ tends to infinity, assuming, for any 
finite $n$, that the right hand side limit of the above equation exists. 
Estimating the limit by setting $N$ large ($N=n+5$ in our numerical 
examples), we distinguish immediately the asymptotic character of the 
sequence $\|R^{(n)}\|_{(e,\delta,\xi)}$: one has that 
$\|R^{(n)}\|_{(e,\delta,\xi)}$ is a decreasing function of $n$ up to 
an {\it optimal normalization order} $n_{opt}$, defined by
\begin{equation}\label{nopt}
\|R^{(n_{opt})}\|_{(e,\delta,\xi)}< \|R^{(n)}\|_{(e,\delta,\xi)}~~~
\mbox{both for $n<n_{opt}$ and $n>n_{opt}$}~~.
\end{equation}
Thus, $n_{opt}=7$ in the left panel of Fig.\ref{fig:ASYM}. As shown 
in the right panel in the same figure, our particular book-keeping 
rule introduced for the detuning parameter is consistent with the 
expected behavior for asymptotic series: $n_{opt}$ is a decreasing 
function of $\delta$. We find the power-law estimate $n_{opt}\sim\delta^{-b}$, 
with $b\approx 1$, while, as a consequence, $\|R^{(n_{opt})}\|_{(e,\delta,\xi)}$ 
increases as $\delta$ increases. 

\begin{figure}
\centering
\includegraphics[width=0.45\textwidth]{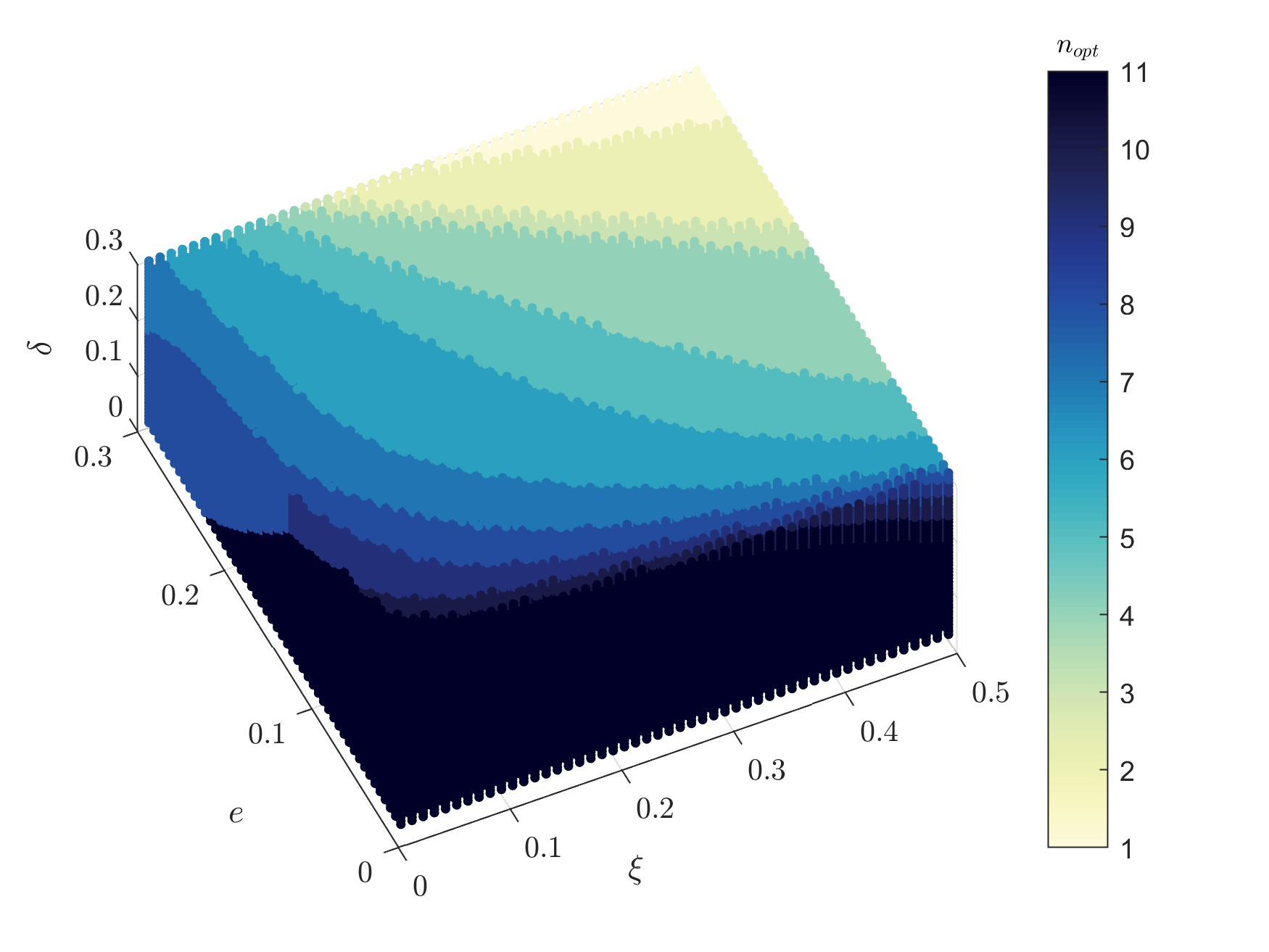}
\includegraphics[width=0.45\textwidth]{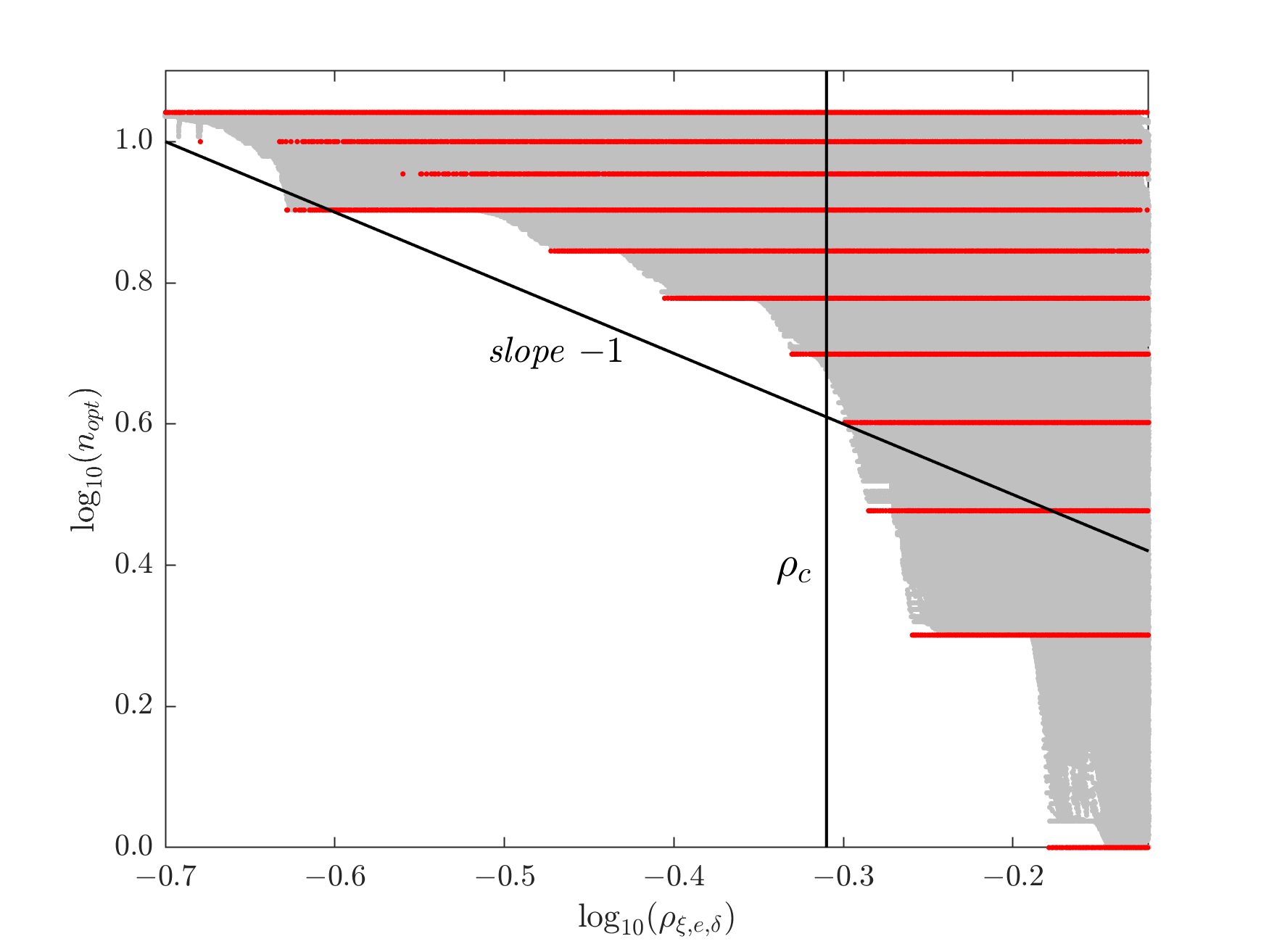}
\caption{For the case of the 1:1 secondary resonance we provide
in the left panel the optimal normalisation order $n$
in the 3-dimensional space of the parameters ($\xi,e,\delta$). 
In right panel the normalisation order $n$ is shown for different
values of the `distance' from the origin $\rho$.}
    \label{fig:CUBEP11S11}
\end{figure}

\begin{figure}
\centering
\includegraphics[width=0.45\textwidth]{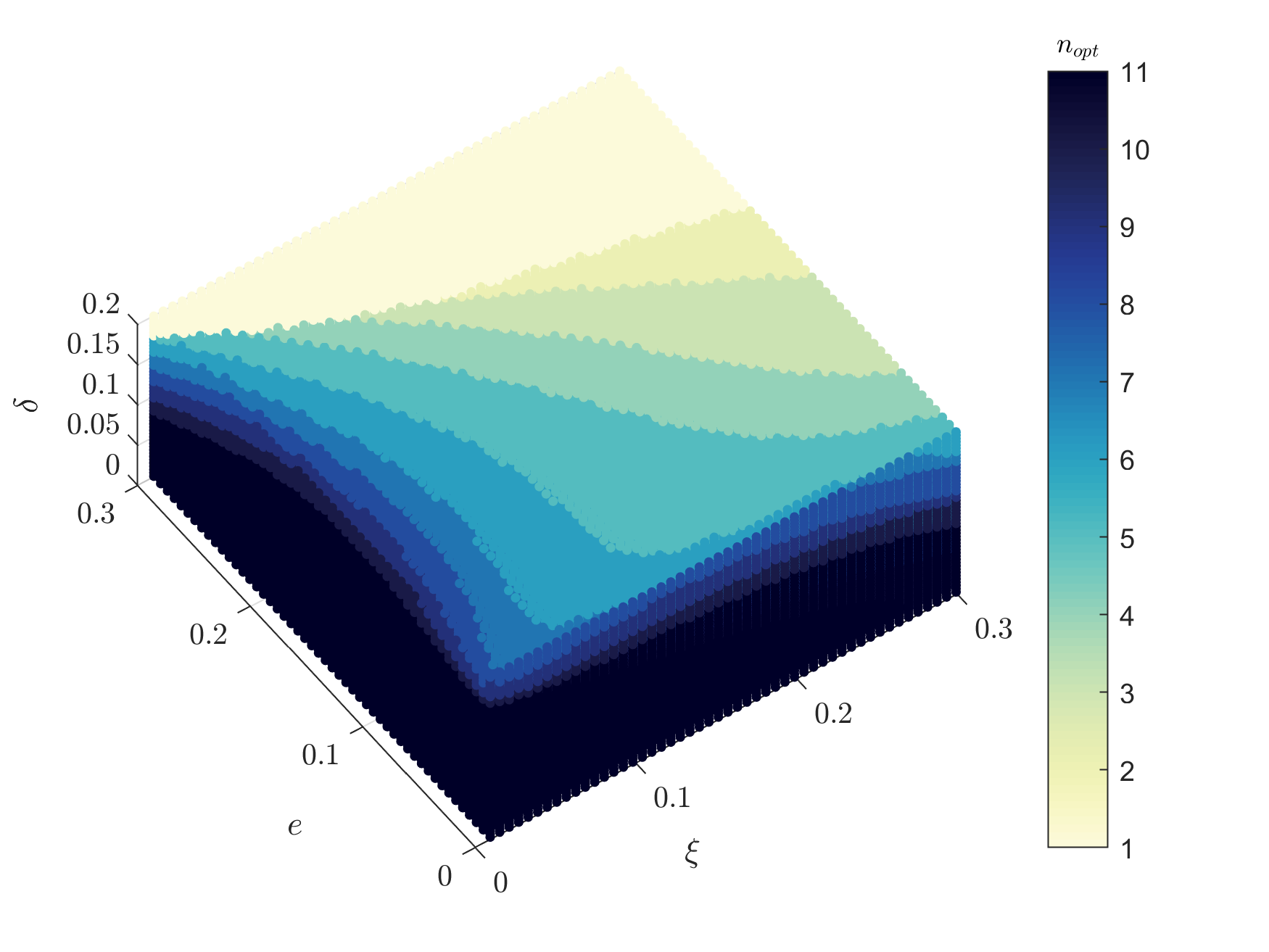}
\includegraphics[width=0.45\textwidth]{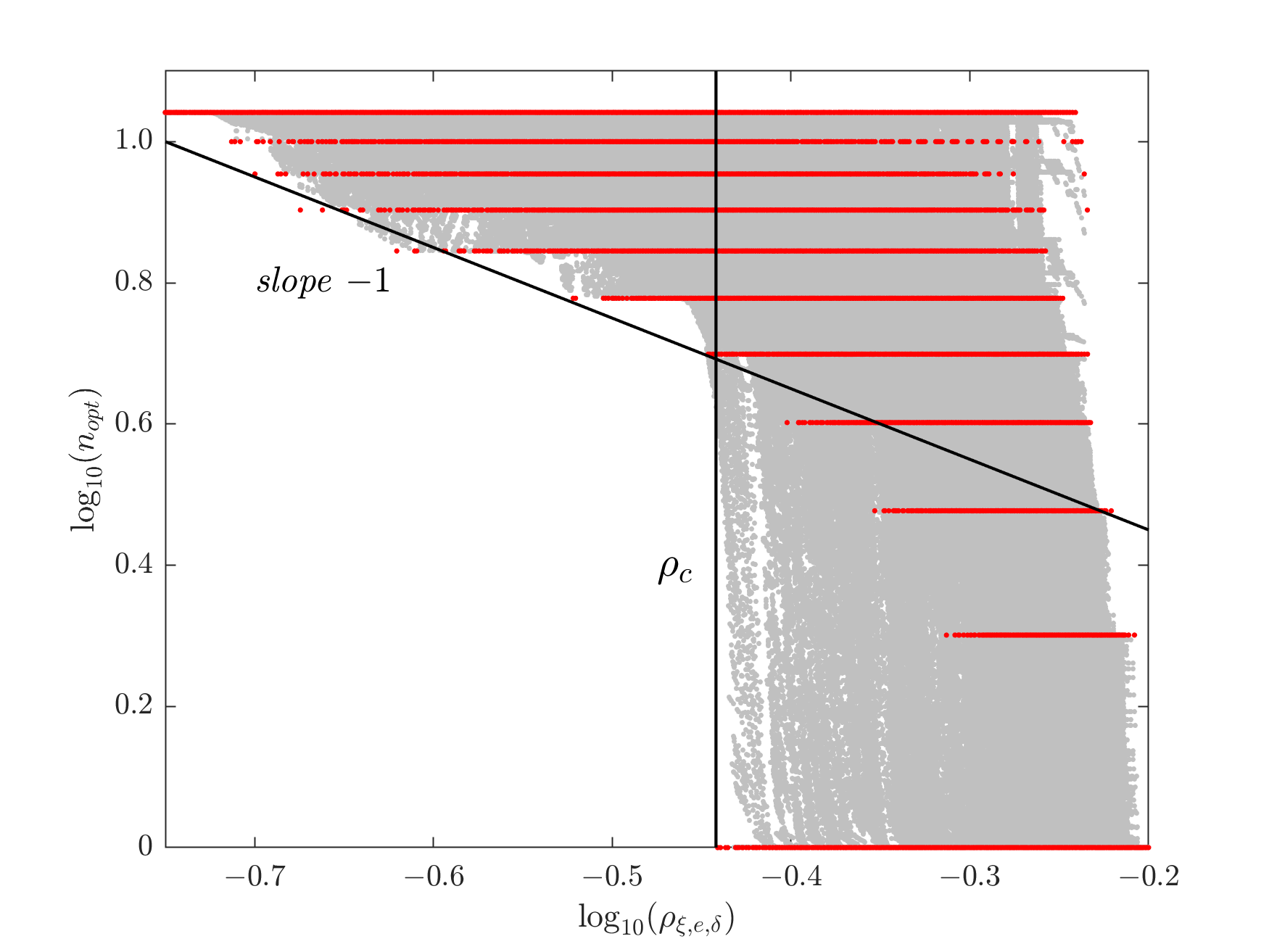}
\caption{Same as in Fig.~\ref{fig:CUBEP11S11} but for the case of the 2:1 secondary resonance.}
    \label{fig:CUBEP11S21}
\end{figure}

\begin{figure}
\centering
\includegraphics[width=0.45\textwidth]{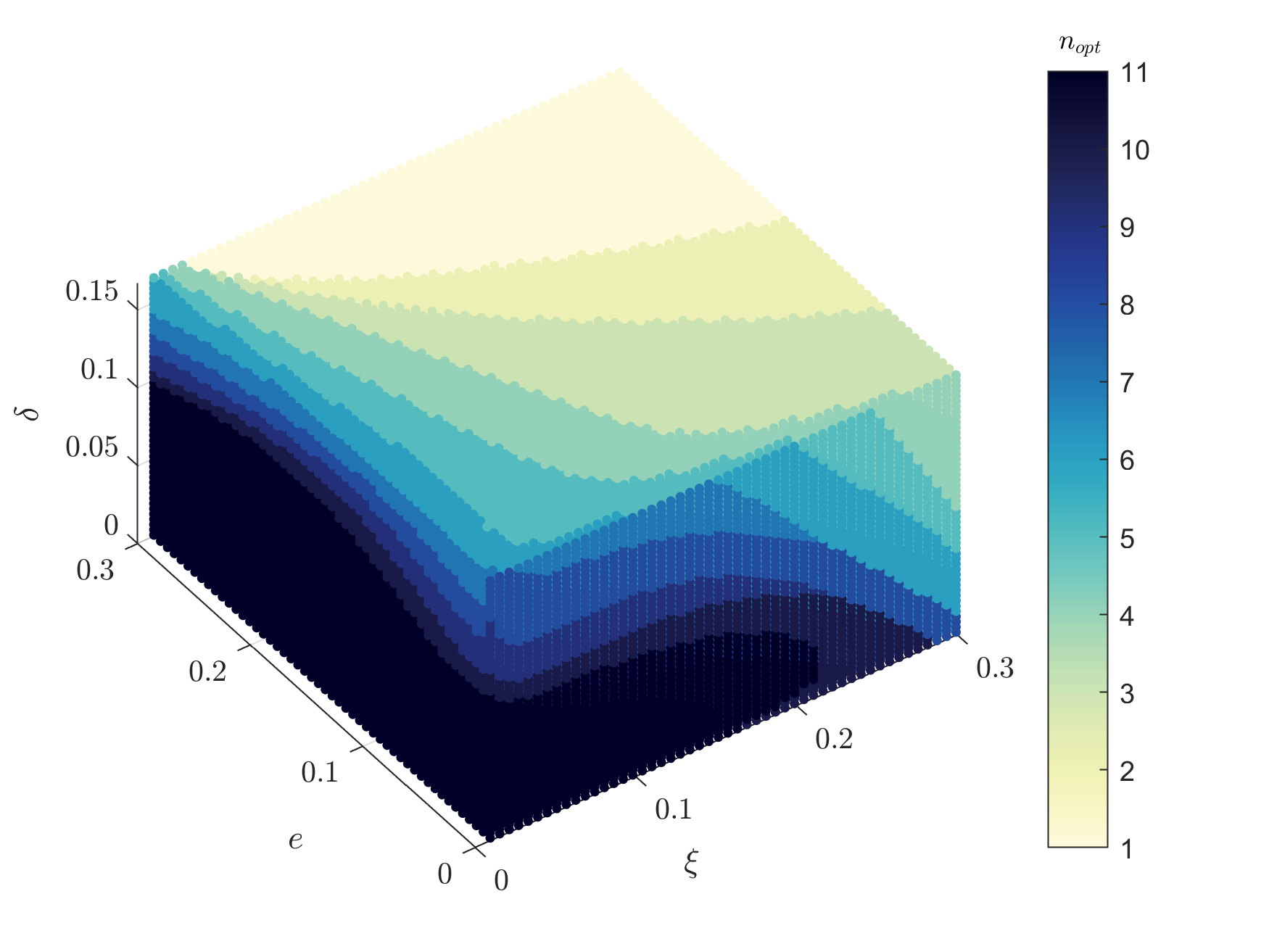}
\includegraphics[width=0.45\textwidth]{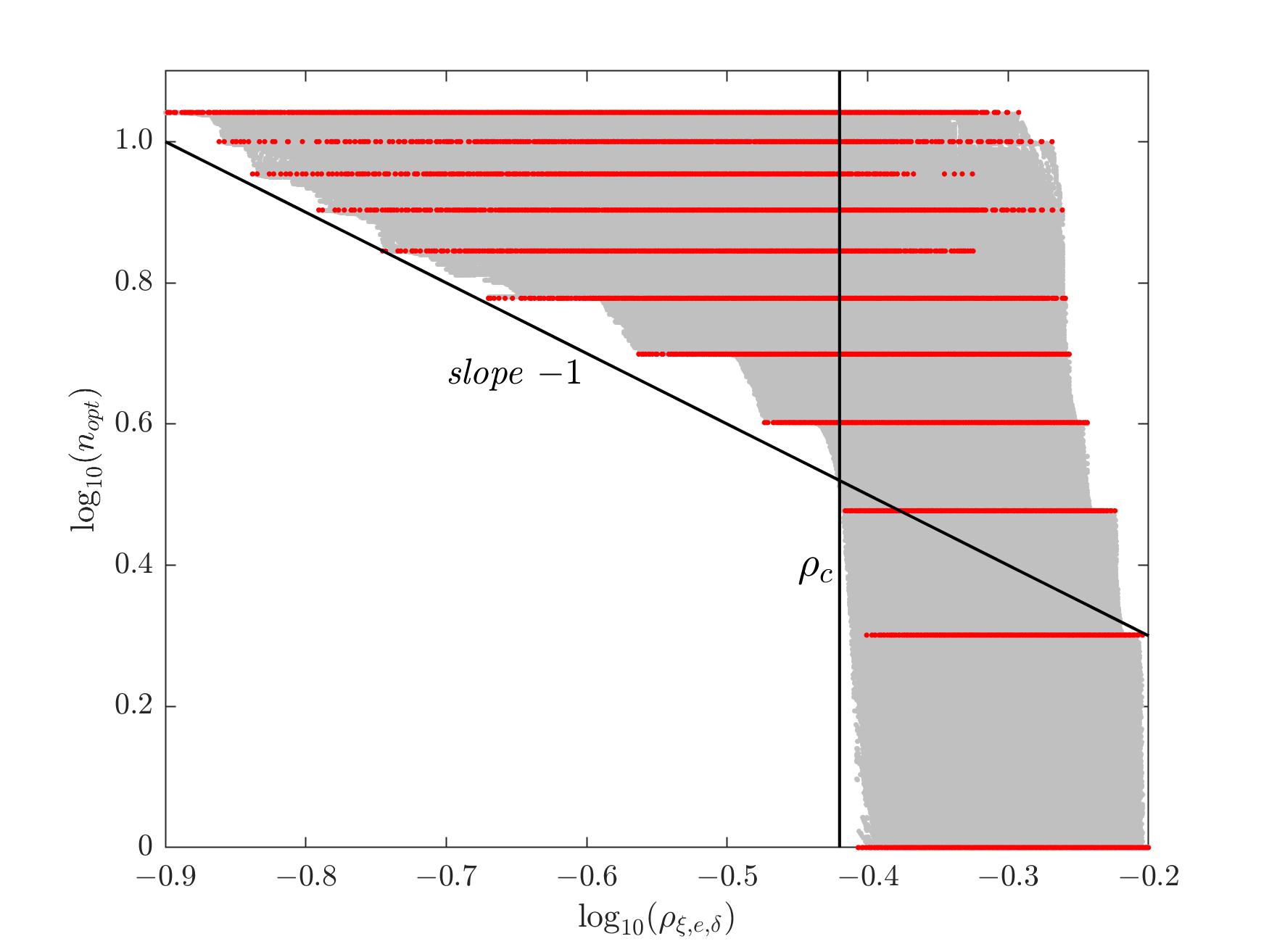}
\caption{Same as in Fig.~\ref{fig:CUBEP11S11} but for the case of the 3:1 secondary resonance.}
    \label{fig:CUBEP11S31}
\end{figure}

Figures \ref{fig:CUBEP11S11}, \ref{fig:CUBEP11S21}, and \ref{fig:CUBEP11S31}
summarize the information on the optimal normalization 
order, estimated by $||R^{(n_{opt})}||_{e,\delta,\xi}$, as a function of the 
three small parameters $e,\delta,\xi$, for the secondary resonances 1:1, 
2:1 and 3:1 respectively. All three figures have a similar structure, 
which represents the general trend expected for asymptotic series, namely 
the fact that the optimal order decreases as the value of the small parameter(s) 
increases. Regarding more precise quantitative estimates, the right panels in 
Figs. 8, 9 and 10 show the dependence of the computed optimal orders $n_{opt}$ 
on a unique quantity $\rho$ representing the `distance' from the origin in 
parameter or phase space, defined as:
\begin{equation}\label{rho}
 \rho = \sqrt{\xi + \delta^2 + e^2}.
\end{equation}
Note that in above expression $\xi$ appears in the first power in the square 
root, since $\xi$ represents a limit in the action space ($J<\xi$ in the 
norm definition; see Eq.~\eqref{eq:remnormtrunc}), thus it represents already the square of the 
distance from the origin in the Poincar\'{e} variables $(X,Y)$. As shown in 
the right panels of Figs.~ 8 to 10, for various combinations of the three 
parameters $(\xi,\delta,e)$ yielding a fixed $\rho$ below some threshold 
$\rho<\rho_c$, one obtains various optimal orders bounded from below 
according to $n_{opt}\geq n_{opt,min}$. The lack of upper limit in the 
optimal order simply reflects the integrability of the model when $e=0$ 
(a fact which implies that the series are convergent in this case for 
appropriate bounds in $\xi$ and $\delta$). On the other hand, the lower 
bound is close to the power law $n_{opt,min}\propto \rho^{-1}$, a 
relation which is characteristic of resonant normal forms (see 
\cite{Efthym2004} for more details). This power-law behavior breaks, however, 
at $\rho\approx\rho_c$. The behavior of the series there is dominated again 
by its dependence on the eccentricity: we find that, independently of the 
asphericity value, chaos prevails in phase space when the eccentricity 
acquires values around $e\approx 0.2-0.3$. This fact is connected with 
the resonance overlap between the 1:1 and 3:2 {\it primary} resonances. 
A rough application of Chirikov's resonance overlap criterion shows that 
this happens at eccentricities $e_c \geqslant 2/7$, a value which marks the onset 
of large chaos and the collapse of the integrable representation of the 
system by the normal form approach. 

These results are verified also in Figs. 11 to 13, which show the dependence 
of the optimal normalization order, as well as the optimal remainder value 
(i.e. the error at the optimal order) as a function of the detuning and 
orbital parameters $\delta$ and $e$, for three different values of $\xi$, 
namely $\xi=0.01$, $0.1$ and $\xi_{max}$, with $\xi_{max}=0.5$ in the 
case of the 1:1 secondary resonance, while $\xi_{max}=0.3$ for the 2:1 
and 3:1 secondary resonances. The value of $\xi_{max}$ represents the extend 
of the regular domain up to about the separatrix limit of the primary 
resonance (compare with the phase portraits in Figs.~\ref{fig:PSSP11S11},
\ref{fig:PSSP11S21} and \ref{fig:PSSP11S31}), while the  
value $\xi=0.01$ represents a domain within which most bifurcations 
take place. The value $\xi=0.1$ is intermediate between the two previous 
cases. Besides observing the general collapse of the normal form approach 
close to the separatrix limit of the primary resonance, where chaos 
prevails, one notices also the nearly uniform, in the rest of the 
parameters, collapse of the normal form approach when the eccentricity 
exceeds a value $\approx 0.2 - 0.3$, which represents the threshold to large 
scale chaos due to the resonance overlap between the 1:1 and 3:2 primary 
resonances. At any rate, far from these limits one obtains a remarkably 
good behavior of the normal form, with errors around $10^{-10}$ or 
smaller very close to the origin, where most bifurcation phenomena 
take place, and still quite low ($\sim 10^{-5}$) at intermediate 
distances from the origin, both in phase space and in parameter space.


\begin{figure}
\centering
\includegraphics[width=0.9\textwidth]{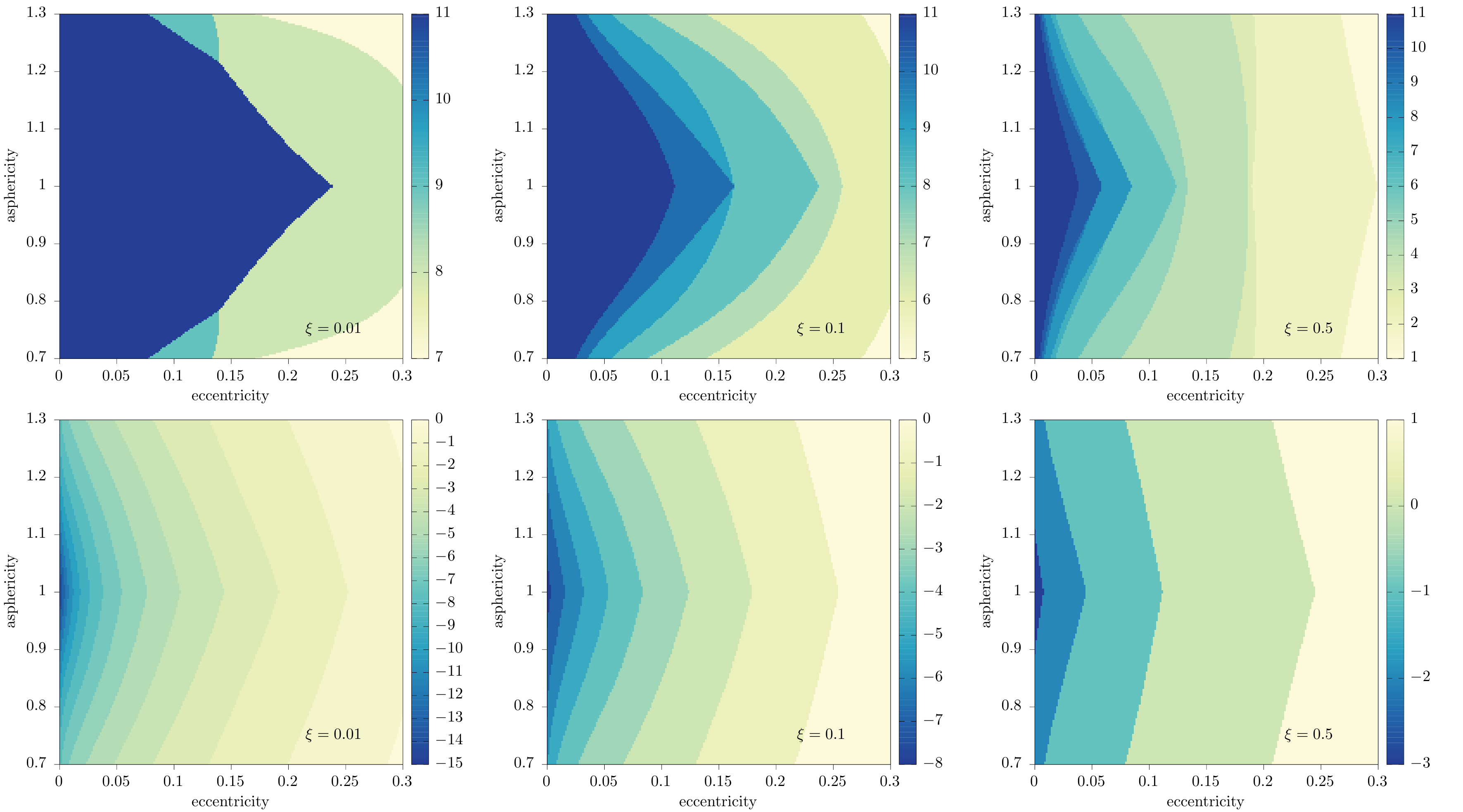}
\caption{The optimal order $n$ (top row) and the error estimates in powers 
of 10 (bottom row) in the $(e,\delta)$ plane for different values of $\xi$
in the case of the 1:1 secondary resonance.}
    \label{fig:P11S11-OPTORDERERROR}
\end{figure}
 
\begin{figure}
\centering
\includegraphics[width=0.9\textwidth]{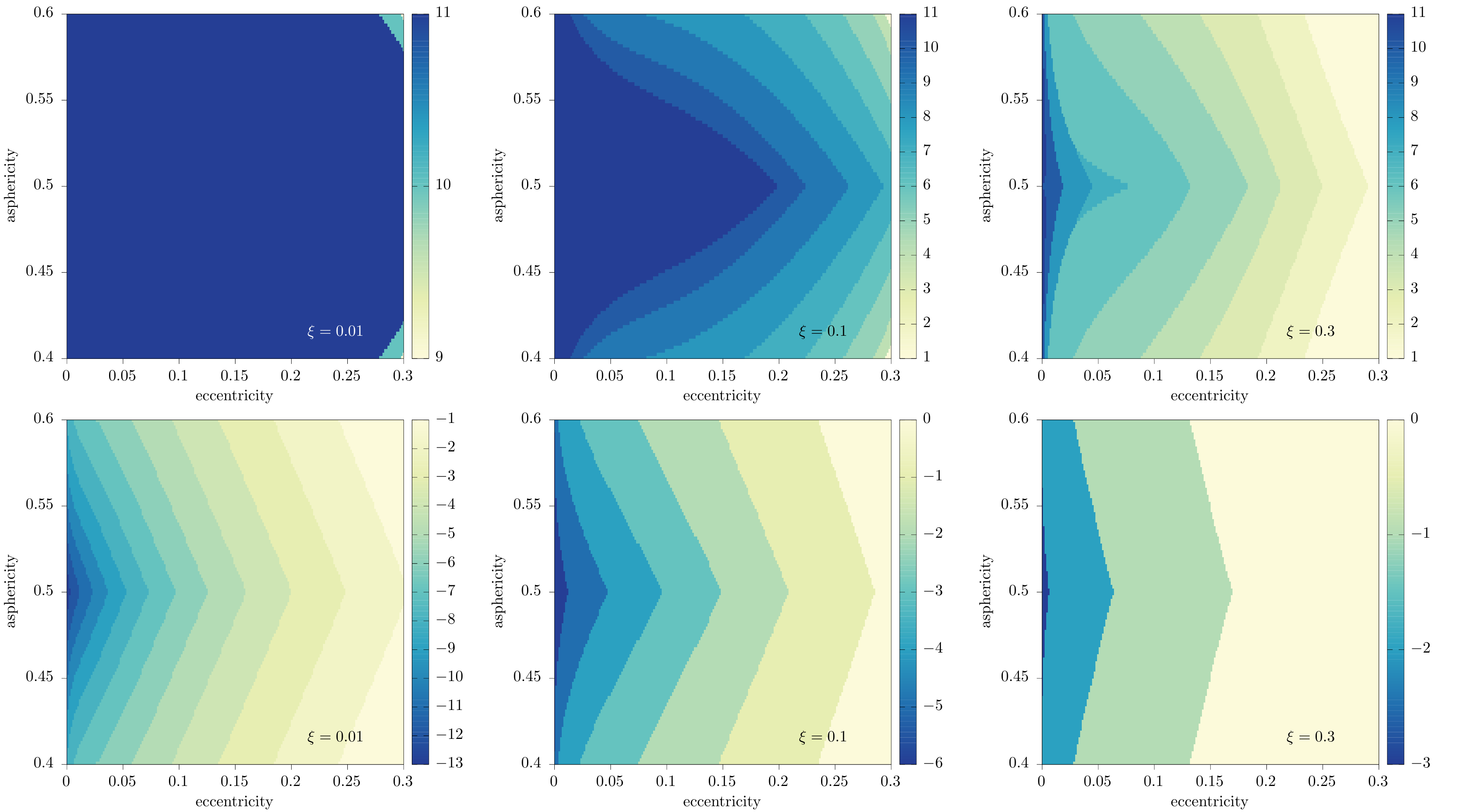}
\caption{Same as in Fig.~\ref{fig:P11S11-OPTORDERERROR} but for the case of the 2:1 secondary resonance.}
    \label{fig:P11S21-OPTORDERERROR}
\end{figure}
 

\begin{figure}
\centering
\includegraphics[width=0.9\textwidth]{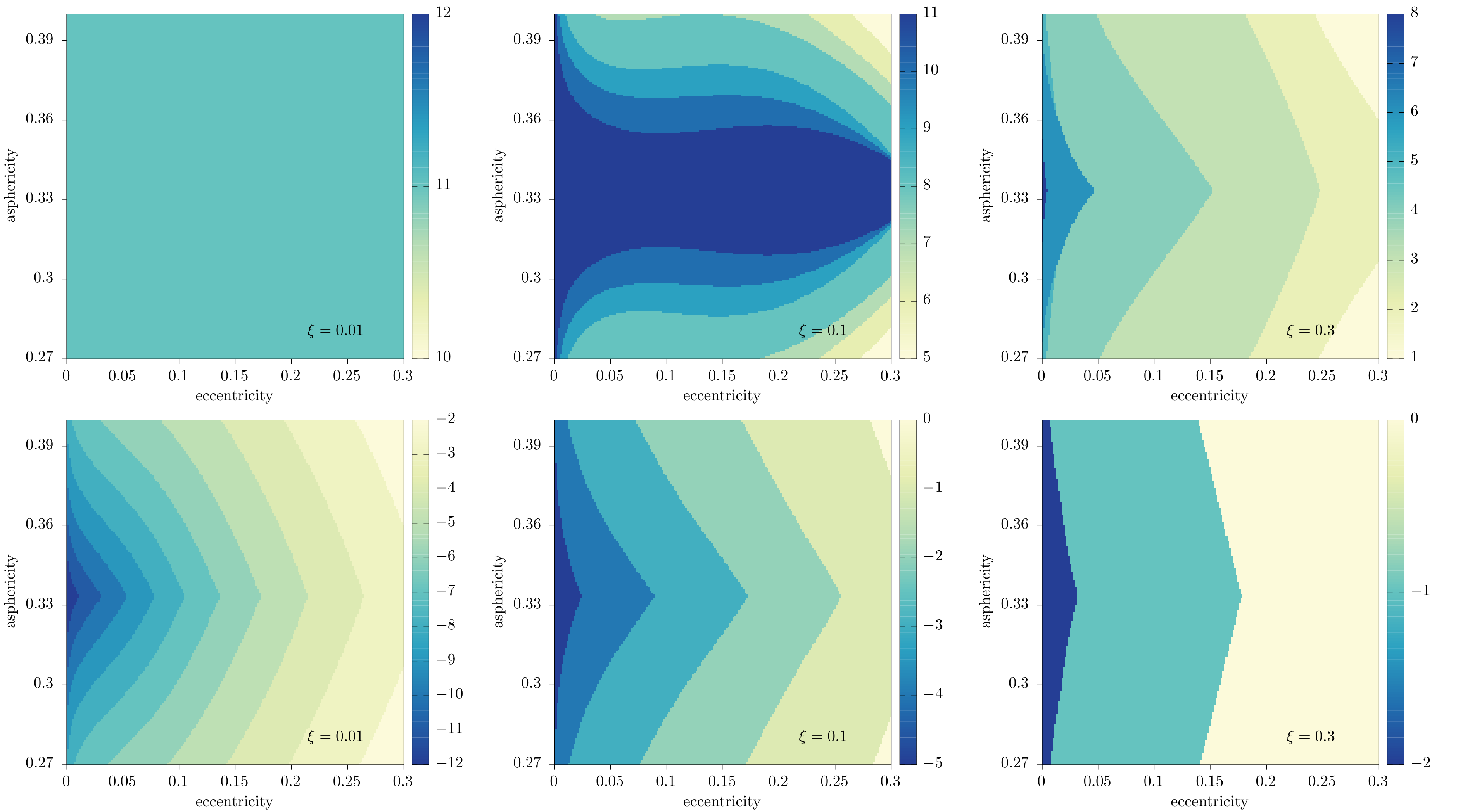}
\caption{Same as in Fig.~\ref{fig:P11S11-OPTORDERERROR} but for the case of the 3:1 secondary resonance.}
    \label{fig:P11S31-OPTORDERERROR}
\end{figure}

\section{Conclusions}\label{sec:conclusions}
The normal form theory can be used to study a wide variety 
of astronomical systems. The study of resonances, primary and 
secondary, can give us very important results in understanding
and exploiting the natural dynamics of the system. In this work,
we further generalise the method presented in \cite{MNRASpaper}
for the study of secondary resonances. The spin-orbit model
still serves as our test problem to apply the proposed 
techniques and study their efficiency. The result for the
2:1 secondary resonance are here extended and additional
secondary resonances are studied (1:1, 3:1), confirming that
our method is generally applicable.

The proposed canonical normalisation scheme, when applied to 
each particular secondary resonance, allows us to compute
an integrable approximation that describes accurately the 
dynamics in the domain of ordered motion. The derived 
expressions result in polynomial functions in Poincar\'e
variables, which allow us to retrieve useful information
for the system in a broad range of the parameter space
($e,\ep$). Moreover, back-transforming the 
propagated 1 D.O.F. dynamics to the original variables,
allow us to obtain accurately the time evolution of the
satellite's spin.

The difference with other normalisation methods proposed 
in the literature, is the exploitation of the detuning 
and book-keeping techniques
to design a normalisation scheme that is efficient, robust 
and algorithmically convenient. The detailed analysis of the 
error behavior in our normal form constructions shows that, 
even with this non-classical choice of term ordering
in the Hamiltonian function, the asymptotic behavior of the 
remainder still remains. Our constructions, not only 
accurately depict the phase-space in parameter space
about the secondary resonances, but also cover with sufficient 
precision a broad region around them.

From an application point of view, the results presented here 
could be useful in explaining a series of astronomical   
observations related to irregularly shaped satellites and 
moonlets in distant binary systems. The analytically
derived expressions could provide parameter dependent formulas 
for the libration angle of the rotating body. Given such kind of
measurements, those formulas can be used to fit the data and 
provide estimates for the eccentricity of the satellite's 
orbits, as well as the size of its equatorial bulge.

The success of the method in this basic setup, motivates
us to pursue future applications in any type of secondary resonances 
in orbital and rotational motion of astronomical objects.
In addition, further adaptation of the technique to work with 
more detailed models of the spin-orbit coupling is feasible.
Successful implementations in other cases will solidify the
method as a useful tool for the general studies of resonant
phenomena.

\vglue1cm

{\bf Acknowledgements.} 
AC was supported by GNFM-INdAM and acknowledges MIUR Excellence Department Project awarded to the Department of Mathematics of the University of Rome Tor Vergata (CUP E83C18000100006). GP was supported by GNFM-INdAM.



\appendix
\section{Appendix}

The $n$-th order normal form of Eq. (\ref{HXY11}), expressed 
in Poincar\'e variables, has the form
$$
H=Z^{(n)}=\sum_{i_1,i_2,i_3,i_4,i_5} a_{i_1,i_2,i_3,i_4,i_5}
\lambda^{i_1} e^{i_2} \delta^{i_3} X^{i_4} Y^{i_5}
$$
with exponents $i_j$, $j=1,\ldots,6$, determined by the book-keeping rules. 
The entire form of $Z^{(6)}$ is given in the
following tables for all low-order secondary resonances of the
1:1 primary resonance: 
Table~\ref{tab:z6p11s11} contains the resonant construction for the 1:1 
secondary resonance, Table~\ref{tab:z6p11s21} for the 2:1 
secondary resonance and Table~\ref{tab:z6p11s31} for the 3:1 
secondary resonance.
\tiny{
\begin{table}   
\setlength{\tabcolsep}{1pt}
\caption{Coefficient list for the 6th order normal form in Poincar\'e 
variables for the 1:1 secondary resonance of the 1:1 primary spin-orbit resonance.}
\begin{minipage}[t]{.3\columnwidth}
    \begin{tabular}[t]{cccccc} 
        \hline
        $i_1$ & $i_2$ & $i_3$ & $i_4$ & $i_5$ & coeff \\
        \hline
1	&	0	&	1	&	2	&	0	&	1/2	\\
1	&	0	&	1	&	2	&	0	&	1/2	\\
1	&	0	&	1	&	0	&	2	&	1/2	\\
2	&	2	&	0	&	2	&	0	&	-1/2	\\
2	&	0	&	0	&	4	&	0	&	-1/16	\\
2	&	1	&	0	&	2	&	1	&	1/2	\\
2	&	2	&	0	&	0	&	2	&	-3/2	\\
2	&	0	&	0	&	2	&	2	&	-1/8	\\
2	&	1	&	0	&	0	&	3	&	1/2	\\
2	&	0	&	0	&	0	&	4	&	-1/16	\\
3	&	2	&	0	&	2	&	0	&	1/4	\\
3	&	1	&	0	&	2	&	1	&	-1/8	\\
3	&	2	&	0	&	0	&	2	&	3/4	\\
3	&	1	&	0	&	0	&	3	&	-1/8	\\
3	&	2	&	1	&	2	&	0	&	-1/4	\\
3	&	1	&	1	&	2	&	1	&	1/4	\\
3	&	2	&	1	&	0	&	2	&	-5/4	\\
3	&	1	&	1	&	0	&	3	&	1/4	\\
4	&	2	&	0	&	2	&	0	&	-1/32	\\
4	&	4	&	0	&	2	&	0	&	-115/192	\\
4	&	2	&	0	&	4	&	0	&	-9/32	\\
4	&	0	&	0	&	6	&	0	&	-1/128	\\
4	&	3	&	0	&	2	&	1	&	59/48	\\
4	&	1	&	0	&	4	&	1	&	3/32	\\
4	&	2	&	0	&	0	&	2	&	-3/32	\\
4	&	4	&	0	&	0	&	2	&	-275/192	\\
4	&	2	&	0	&	2	&	2	&	-3/4	\\
4	&	0	&	0	&	4	&	2	&	-3/128	\\
4	&	3	&	0	&	0	&	3	&	169/144	\\
4	&	1	&	0	&	2	&	3	&	3/16	\\
4	&	2	&	0	&	0	&	4	&	-15/32	\\
4	&	0	&	0	&	2	&	4	&	-3/128	\\
4	&	1	&	0	&	0	&	5	&	3/32	\\
4	&	0	&	0	&	0	&	6	&	-1/128	\\
4	&	2	&	1	&	2	&	0	&	7/16	\\
4	&	1	&	1	&	2	&	1	&	-5/32	\\
4	&	2	&	1	&	0	&	2	&	21/16	\\
4	&	1	&	1	&	0	&	3	&	-5/32	\\
4	&	2	&	2	&	2	&	0	&	25/48	\\
4	&	1	&	2	&	2	&	1	&	-1/16	\\
4	&	2	&	2	&	0	&	2	&	-53/48	\\
4	&	1	&	2	&	0	&	3	&	-1/16	\\
5	&	4	&	0	&	2	&	0	&	757/384	\\
5	&	2	&	0	&	4	&	0	&	15/256	\\
5	&	3	&	0	&	2	&	1	&	-673/384	\\
5	&	1	&	0	&	4	&	1	&	-15/512	\\
5	&	4	&	0	&	0	&	2	&	1607/384	\\
5	&	2	&	0	&	2	&	2	&	45/128	\\
5	&	3	&	0	&	0	&	3	&	-1757/1152	\\
5	&	1	&	0	&	2	&	3	&	-15/256	\\
5	&	2	&	0	&	0	&	4	&	75/256	\\
  		\hline
    \end{tabular}
\end{minipage}
\begin{minipage}[t]{.3\columnwidth}
    \begin{tabular}[t]{cccccc} 
        \hline
        $i_1$ & $i_2$ & $i_3$ & $i_4$ & $i_5$ & coeff \\
        \hline
5	&	1	&	0	&	0	&	5	&	-15/512	\\
5	&	2	&	1	&	2	&	0	&	-11/64	\\
5	&	4	&	1	&	2	&	0	&	-175/96	\\
5	&	2	&	1	&	4	&	0	&	1/64	\\
5	&	0	&	1	&	6	&	0	&	1/128	\\
5	&	3	&	1	&	2	&	1	&	265/384	\\
5	&	1	&	1	&	4	&	1	&	-9/256	\\
5	&	2	&	1	&	0	&	2	&	-17/64	\\
5	&	4	&	1	&	0	&	2	&	461/96	\\
5	&	2	&	1	&	2	&	2	&	31/64	\\
5	&	0	&	1	&	4	&	2	&	3/128	\\
5	&	3	&	1	&	0	&	3	&	-2765/1152	\\
5	&	1	&	1	&	2	&	3	&	-9/128	\\
5	&	2	&	1	&	0	&	4	&	15/32	\\
5	&	0	&	1	&	2	&	4	&	3/128	\\
5	&	1	&	1	&	0	&	5	&	-9/256	\\
5	&	0	&	1	&	0	&	6	&	1/128	\\
5	&	2	&	2	&	2	&	0	&	1/16	\\
5	&	1	&	2	&	2	&	1	&	5/64	\\
5	&	2	&	2	&	0	&	2	&	-1/16	\\
5	&	1	&	2	&	0	&	3	&	5/64	\\
5	&	2	&	3	&	2	&	0	&	23/72	\\
5	&	1	&	3	&	2	&	1	&	1/16	\\
5	&	2	&	3	&	0	&	2	&	59/72	\\
5	&	1	&	3	&	0	&	3	&	1/16	\\
6	&	4	&	0	&	2	&	0	&	17/768	\\
6	&	6	&	0	&	2	&	0	&	-221/1152	\\
6	&	2	&	0	&	4	&	0	&	-113/2048	\\
6	&	4	&	0	&	4	&	0	&	-20891/15360	\\
6	&	2	&	0	&	6	&	0	&	-41/240	\\
6	&	0	&	0	&	8	&	0	&	-5/2048	\\
6	&	3	&	0	&	2	&	1	&	105/256	\\
6	&	5	&	0	&	2	&	1	&	9337/1536	\\
6	&	3	&	0	&	4	&	1	&	12607/7680	\\
6	&	1	&	0	&	6	&	1	&	5/128	\\
6	&	4	&	0	&	0	&	2	&	-363/256	\\
6	&	6	&	0	&	0	&	2	&	-12545/1152	\\
6	&	2	&	0	&	2	&	2	&	-105/1024	\\
6	&	4	&	0	&	2	&	2	&	-19517/2560	\\
6	&	2	&	0	&	4	&	2	&	-219/320	\\
6	&	0	&	0	&	6	&	2	&	-5/512	\\
6	&	3	&	0	&	0	&	3	&	97/256	\\
6	&	5	&	0	&	0	&	3	&	56143/4608	\\
6	&	3	&	0	&	2	&	3	&	13477/3840	\\
6	&	1	&	0	&	4	&	3	&	15/128	\\
6	&	2	&	0	&	0	&	4	&	-97/2048	\\
6	&	4	&	0	&	0	&	4	&	-97051/15360	\\
6	&	2	&	0	&	2	&	4	&	-137/160	\\
6	&	0	&	0	&	4	&	4	&	-15/1024	\\
6	&	3	&	0	&	0	&	5	&	14347/7680	\\
6	&	1	&	0	&	2	&	5	&	15/128	\\	
		\hline
    \end{tabular}
\end{minipage}\hspace{0.3cm}
\begin{minipage}[t]{.3\columnwidth}
    \begin{tabular}[t]{cccccc} 
        \hline
        $i_1$ & $i_2$ & $i_3$ & $i_4$ & $i_5$ & coeff \\
        \hline
6	&	2	&	0	&	0	&	6	&	-329/960	\\
6	&	0	&	0	&	2	&	6	&	-5/512	\\
6	&	1	&	0	&	0	&	7	&	5/128	\\
6	&	0	&	0	&	0	&	8	&	-5/2048	\\
6	&	4	&	1	&	2	&	0	&	4435/2304	\\
6	&	2	&	1	&	4	&	0	&	31/128	\\
6	&	3	&	1	&	2	&	1	&	-2431/768	\\
6	&	1	&	1	&	4	&	1	&	-11/1024	\\
6	&	4	&	1	&	0	&	2	&	8809/2304	\\
6	&	2	&	1	&	2	&	2	&	69/128	\\
6	&	3	&	1	&	0	&	3	&	-375/256	\\
6	&	1	&	1	&	2	&	3	&	-11/512	\\
6	&	2	&	1	&	0	&	4	&	19/64	\\
6	&	1	&	1	&	0	&	5	&	-11/1024	\\
6	&	2	&	2	&	2	&	0	&	-69/512	\\
6	&	4	&	2	&	2	&	0	&	-3803/1152	\\
6	&	2	&	2	&	4	&	0	&	-6065/4608	\\
6	&	0	&	2	&	6	&	0	&	-1/128	\\
6	&	3	&	2	&	2	&	1	&	821/144	\\
6	&	1	&	2	&	4	&	1	&	1/256	\\
6	&	2	&	2	&	0	&	2	&	-173/1536	\\
6	&	4	&	2	&	0	&	2	&	-2051/1152	\\
6	&	2	&	2	&	2	&	2	&	-405/256	\\
6	&	0	&	2	&	4	&	2	&	-3/128	\\
6	&	3	&	2	&	0	&	3	&	23/24	\\
6	&	1	&	2	&	2	&	3	&	1/128	\\
6	&	2	&	2	&	0	&	4	&	-1225/4608	\\
6	&	0	&	2	&	2	&	4	&	-3/128	\\
6	&	1	&	2	&	0	&	5	&	1/256	\\
6	&	0	&	2	&	0	&	6	&	-1/128	\\
6	&	2	&	3	&	2	&	0	&	-89/768	\\
6	&	1	&	3	&	2	&	1	&	-25/1536	\\
6	&	2	&	3	&	0	&	2	&	-107/768	\\
6	&	1	&	3	&	0	&	3	&	-25/1536	\\
6	&	2	&	4	&	2	&	0	&	-871/540	\\
6	&	1	&	4	&	2	&	1	&	-47/768	\\
6	&	2	&	4	&	0	&	2	&	-349/540	\\
6	&	1	&	4	&	0	&	3	&	-47/768	\\
		\hline
    \end{tabular}
\end{minipage}
\label{tab:z6p11s11}
\end{table}
}

\tiny{
\begin{table}
\setlength{\tabcolsep}{1pt}
    \centering
\caption{Coefficient list for the 6th order normal form in Poincar\'e 
variables for the 2:1 secondary resonance of the 1:1 primary spin-orbit resonance.}
\begin{minipage}[t]{.3\columnwidth}
    \begin{tabular}[t]{cccccc} 
        \hline
        $i_1$ & $i_2$ & $i_3$ & $i_4$ & $i_5$ & coeff \\
        \hline
		1	&	1	&	0	&	2	&	0	&	3/16	\\
1	&	1	&	0	&	0	&	2	&	-3/16	\\
1	&	0	&	1	&	2	&	0	&	1/2	\\
1	&	0	&	1	&	0	&	2	&	1/2	\\
2	&	2	&	0	&	2	&	0	&	-89/256	\\
2	&	0	&	0	&	4	&	0	&	-1/16	\\
2	&	2	&	0	&	0	&	2	&	-89/256	\\
2	&	0	&	0	&	2	&	2	&	-1/8	\\
2	&	0	&	0	&	0	&	4	&	-1/16	\\
2	&	1	&	1	&	2	&	0	&	3/8	\\
2	&	1	&	1	&	0	&	2	&	-3/8	\\
3	&	2	&	0	&	2	&	0	&	-1/3	\\
3	&	3	&	0	&	2	&	0	&	-365/4096	\\
3	&	1	&	0	&	4	&	0	&	-3/64	\\
3	&	2	&	0	&	0	&	2	&	-1/3	\\
3	&	3	&	0	&	0	&	2	&	365/4096	\\
3	&	1	&	0	&	0	&	4	&	3/64	\\
3	&	2	&	1	&	2	&	0	&	-223/256	\\
3	&	2	&	1	&	0	&	2	&	-223/256	\\
3	&	1	&	2	&	2	&	0	&	-3/8	\\
3	&	1	&	2	&	0	&	2	&	3/8	\\
4	&	2	&	0	&	2	&	0	&	-1/18	\\
4	&	3	&	0	&	2	&	0	&	-31/240	\\
4	&	4	&	0	&	2	&	0	&	62221/196608	\\
4	&	2	&	0	&	4	&	0	&	3619/10240	\\
4	&	0	&	0	&	6	&	0	&	-1/64	\\
4	&	2	&	0	&	0	&	2	&	-1/18	\\
4	&	3	&	0	&	0	&	2	&	31/240	\\
4	&	4	&	0	&	0	&	2	&	62221/196608	\\
4	&	2	&	0	&	2	&	2	&	-2981/5120	\\
4	&	0	&	0	&	4	&	2	&	-3/64	\\
4	&	2	&	0	&	0	&	4	&	3619/10240	\\
4	&	0	&	0	&	2	&	4	&	-3/64	\\
4	&	0	&	0	&	0	&	6	&	-1/64	\\
4	&	2	&	1	&	2	&	0	&	-22/9	\\
4	&	3	&	1	&	2	&	0	&	239/256	\\
4	&	1	&	1	&	4	&	0	&	15/64	\\
4	&	2	&	1	&	0	&	2	&	-22/9	\\
4	&	3	&	1	&	0	&	2	&	-239/256	\\
4	&	1	&	1	&	0	&	4	&	-15/64	\\
4	&	2	&	2	&	2	&	0	&	-3/256	\\
4	&	2	&	2	&	0	&	2	&	-3/256	\\
4	&	1	&	3	&	2	&	0	&	13/16	\\
4	&	1	&	3	&	0	&	2	&	-13/16	\\
5	&	3	&	0	&	2	&	0	&	-3/160	\\
5	&	4	&	0	&	2	&	0	&	186527/115200	\\
5	&	5	&	0	&	2	&	0	&	-458595/1048576	\\
5	&	2	&	0	&	4	&	0	&	23/120	\\
5	&	3	&	0	&	4	&	0	&	-15443/409600	\\
5	&	1	&	0	&	6	&	0	&	-57/1024	\\
        \hline
    \end{tabular}
\end{minipage}
\begin{minipage}[t]{.3\columnwidth}
    \begin{tabular}[t]{cccccc} 
        \hline
        $i_1$ & $i_2$ & $i_3$ & $i_4$ & $i_5$ & coeff \\
        \hline
5	&	3	&	0	&	0	&	2	&	3/160	\\
5	&	4	&	0	&	0	&	2	&	186527/115200	\\
5	&	5	&	0	&	0	&	2	&	458595/1048576	\\
5	&	2	&	0	&	2	&	2	&	-37/60	\\
5	&	1	&	0	&	4	&	2	&	-57/1024	\\
5	&	2	&	0	&	0	&	4	&	23/120	\\
5	&	3	&	0	&	0	&	4	&	15443/409600	\\
5	&	1	&	0	&	2	&	4	&	57/1024	\\
5	&	1	&	0	&	0	&	6	&	57/1024	\\
5	&	2	&	1	&	2	&	0	&	-19/27	\\
5	&	3	&	1	&	2	&	0	&	2213/14400	\\
5	&	4	&	1	&	2	&	0	&	1033099/589824	\\
5	&	2	&	1	&	4	&	0	&	36991/19200	\\
5	&	0	&	1	&	6	&	0	&	1/32	\\
5	&	2	&	1	&	0	&	2	&	-19/27	\\
5	&	3	&	1	&	0	&	2	&	-2213/14400	\\
5	&	4	&	1	&	0	&	2	&	1033099/589824	\\
5	&	2	&	1	&	2	&	2	&	73457/19200	\\
5	&	0	&	1	&	4	&	2	&	3/32	\\
5	&	2	&	1	&	0	&	4	&	36991/19200	\\
5	&	0	&	1	&	2	&	4	&	3/32	\\
5	&	0	&	1	&	0	&	6	&	1/32	\\
5	&	2	&	2	&	2	&	0	&	-208/27	\\
5	&	3	&	2	&	2	&	0	&	6989/4096	\\
5	&	1	&	2	&	4	&	0	&	-49/128	\\
5	&	2	&	2	&	0	&	2	&	-208/27	\\
5	&	3	&	2	&	0	&	2	&	-6989/4096	\\
5	&	1	&	2	&	0	&	4	&	49/128	\\
5	&	2	&	3	&	2	&	0	&	39/64	\\
5	&	2	&	3	&	0	&	2	&	39/64	\\
5	&	1	&	4	&	2	&	0	&	-15/32	\\
5	&	1	&	4	&	0	&	2	&	15/32	\\
6	&	4	&	0	&	2	&	0	&	42653/57600	\\
6	&	5	&	0	&	2	&	0	&	-10105549/21504000	\\
6	&	6	&	0	&	2	&	0	&	-561054889/1509949440	\\
6	&	2	&	0	&	4	&	0	&	23/720	\\
6	&	3	&	0	&	4	&	0	&	-2723/19200	\\
6	&	4	&	0	&	4	&	0	&	-106116307/32256000	\\
6	&	2	&	0	&	6	&	0	&	-358757/1843200	\\
6	&	0	&	0	&	8	&	0	&	-5/512	\\
6	&	4	&	0	&	0	&	2	&	42653/57600	\\
6	&	5	&	0	&	0	&	2	&	10105549/21504000	\\
6	&	6	&	0	&	0	&	2	&	-561054889/1509949440	\\
6	&	2	&	0	&	2	&	2	&	-37/360	\\
6	&	4	&	0	&	2	&	2	&	-22740677/43008000	\\
6	&	2	&	0	&	4	&	2	&	-761357/614400	\\
6	&	0	&	0	&	6	&	2	&	-5/128	\\
6	&	2	&	0	&	0	&	4	&	23/720	\\
6	&	3	&	0	&	0	&	4	&	2723/19200	\\
6	&	4	&	0	&	0	&	4	&	-106116307/32256000	\\
		\hline
    \end{tabular}
\end{minipage}\hspace{0.3cm}
\begin{minipage}[t]{.3\columnwidth}
    \begin{tabular}[t]{cccccc} 
        \hline
        $i_1$ & $i_2$ & $i_3$ & $i_4$ & $i_5$ & coeff \\
        \hline
6	&	2	&	0	&	2	&	4	&	-761357/614400	\\
6	&	0	&	0	&	4	&	4	&	-15/256	\\
6	&	2	&	0	&	0	&	6	&	-358757/1843200	\\
6	&	0	&	0	&	2	&	6	&	-5/128	\\
6	&	0	&	0	&	0	&	8	&	-5/512	\\
6	&	3	&	1	&	2	&	0	&	-1723/21600	\\
6	&	4	&	1	&	2	&	0	&	14951113/864000	\\
6	&	5	&	1	&	2	&	0	&	-39531961/23592960	\\
6	&	2	&	1	&	4	&	0	&	502/225	\\
6	&	3	&	1	&	4	&	0	&	-7916917/1024000	\\
6	&	1	&	1	&	6	&	0	&	111/512	\\
6	&	3	&	1	&	0	&	2	&	1723/21600	\\
6	&	4	&	1	&	0	&	2	&	14951113/864000	\\
6	&	5	&	1	&	0	&	2	&	39531961/23592960	\\
6	&	2	&	1	&	2	&	2	&	-196/225	\\
6	&	1	&	1	&	4	&	2	&	111/512	\\
6	&	2	&	1	&	0	&	4	&	502/225	\\
6	&	3	&	1	&	0	&	4	&	7916917/1024000	\\
6	&	1	&	1	&	2	&	4	&	-111/512	\\
6	&	1	&	1	&	0	&	6	&	-111/512	\\
6	&	2	&	2	&	2	&	0	&	-112/27	\\
6	&	3	&	2	&	2	&	0	&	108329/18000	\\
6	&	4	&	2	&	2	&	0	&	12143431/4423680	\\
6	&	2	&	2	&	4	&	0	&	-9556079/2304000	\\
6	&	0	&	2	&	6	&	0	&	-1/16	\\
6	&	2	&	2	&	0	&	2	&	-112/27	\\
6	&	3	&	2	&	0	&	2	&	-108329/18000	\\
6	&	4	&	2	&	0	&	2	&	12143431/4423680	\\
6	&	2	&	2	&	2	&	2	&	26200921/1152000	\\
6	&	0	&	2	&	4	&	2	&	-3/16	\\
6	&	2	&	2	&	0	&	4	&	-9556079/2304000	\\
6	&	0	&	2	&	2	&	4	&	-3/16	\\
6	&	0	&	2	&	0	&	6	&	-1/16	\\
6	&	2	&	3	&	2	&	0	&	-1312/81	\\
6	&	3	&	3	&	2	&	0	&	43/10240	\\
6	&	1	&	3	&	4	&	0	&	139/384	\\
6	&	2	&	3	&	0	&	2	&	-1312/81	\\
6	&	3	&	3	&	0	&	2	&	-43/10240	\\
6	&	1	&	3	&	0	&	4	&	-139/384	\\
6	&	2	&	4	&	2	&	0	&	-1179/1280	\\
6	&	2	&	4	&	0	&	2	&	-1179/1280	\\
6	&	1	&	5	&	2	&	0	&	1/5	\\
6	&	1	&	5	&	0	&	2	&	-1/5	\\
		\hline
    \end{tabular}
\end{minipage}
\label{tab:z6p11s21}
\end{table}
}

\tiny{
\begin{table}
\setlength{\tabcolsep}{1pt}
\caption{Coefficient list for the 6th order normal form in Poincar\'e 
variables for the 3:1 secondary resonance of the 1:1 primary spin-orbit resonance.}
\centering
   \begin{minipage}[t]{.3\columnwidth}
    \begin{tabular}[t]{cccccc} 
        \hline
        $i_1$ & $i_2$ & $i_3$ & $i_4$ & $i_5$ & coeff \\
        \hline
		1	&	0	&	1	&	2	&	0	&	1/2	\\
1	&	0	&	1	&	0	&	2	&	1/2	\\
2	&	2	&	0	&	2	&	0	&	-2/15	\\
2	&	0	&	0	&	4	&	0	&	-1/16	\\
2	&	1	&	0	&	2	&	1	&	1/(2 $\sqrt{3}$)	\\
2	&	2	&	0	&	0	&	2	&	-2/15	\\
2	&	0	&	0	&	2	&	2	&	-1/8	\\
2	&	1	&	0	&	0	&	3	&	-1/(6 $\sqrt{3}$)	\\
2	&	0	&	0	&	0	&	4	&	-1/16	\\
3	&	2	&	0	&	2	&	0	&	-1/12	\\
3	&	1	&	0	&	2	&	1	&	1/(16 $\sqrt{3}$)	\\
3	&	2	&	0	&	0	&	2	&	-1/12	\\
3	&	1	&	0	&	0	&	3	&	-1/(48 $\sqrt{3}$)	\\
3	&	2	&	1	&	2	&	0	&	41/100	\\
3	&	1	&	1	&	2	&	1	&	$\sqrt{3}$/4	\\
3	&	2	&	1	&	0	&	2	&	41/100	\\
3	&	1	&	1	&	0	&	3	&	-1/(4 $\sqrt{3}$)	\\
4	&	2	&	0	&	2	&	0	&	-1/192	\\
4	&	4	&	0	&	2	&	0	&	-7837/24000	\\
4	&	2	&	0	&	4	&	0	&	-277/1200	\\
4	&	0	&	0	&	6	&	0	&	-3/128	\\
4	&	3	&	0	&	2	&	1	&	-97/(400 $\sqrt{3}$)	\\
4	&	2	&	0	&	0	&	2	&	-1/192	\\
4	&	4	&	0	&	0	&	2	&	-7837/24000	\\
4	&	2	&	0	&	2	&	2	&	-277/600	\\
4	&	0	&	0	&	4	&	2	&	-9/128	\\
4	&	3	&	0	&	0	&	3	&	97/(1200 $\sqrt{3}$)	\\
4	&	2	&	0	&	0	&	4	&	-277/1200	\\
4	&	0	&	0	&	2	&	4	&	-9/128	\\
4	&	0	&	0	&	0	&	6	&	-3/128	\\
4	&	2	&	1	&	2	&	0	&	-13/16	\\
4	&	1	&	1	&	2	&	1	&	11 $\sqrt{3}$/64	\\
4	&	2	&	1	&	0	&	2	&	-13/16	\\
4	&	1	&	1	&	0	&	3	&	-11/(64 $\sqrt{3}$)	\\
4	&	2	&	2	&	2	&	0	&	7533/1000	\\
4	&	1	&	2	&	2	&	1	&	-3 $\sqrt{3}$/16	\\
4	&	2	&	2	&	0	&	2	&	7533/1000	\\
4	&	1	&	2	&	0	&	3	&	$\sqrt{3}$/16	\\
5	&	4	&	0	&	2	&	0	&	257/4200	\\
5	&	2	&	0	&	4	&	0	&	7/384	\\
5	&	3	&	0	&	2	&	1	&	-1201/(11200 $\sqrt{3}$)	\\
5	&	4	&	0	&	0	&	2	&	257/4200	\\
5	&	2	&	0	&	2	&	2	&	7/192	\\
5	&	3	&	0	&	0	&	3	&	1201/(33600 $\sqrt{3}$)	\\
5	&	2	&	0	&	0	&	4	&	7/384	\\
5	&	2	&	1	&	2	&	0	&	-11/128	\\
5	&	4	&	1	&	2	&	0	&	-194453/40000	\\
5	&	2	&	1	&	4	&	0	&	-66191/16000	\\
5	&	0	&	1	&	6	&	0	&	9/128	\\
5	&	3	&	1	&	2	&	1	&	-14981 $\sqrt{3}$/16000	\\
        \hline
    \end{tabular}
\end{minipage}
\begin{minipage}[t]{.3\columnwidth}
    \begin{tabular}[t]{cccccc} 
        \hline
        $i_1$ & $i_2$ & $i_3$ & $i_4$ & $i_5$ & coeff \\
        \hline
5	&	1	&	1	&	4	&	1	&	27 $\sqrt{3}$/128	\\
5	&	2	&	1	&	0	&	2	&	-11/128	\\
5	&	4	&	1	&	0	&	2	&	-194453/40000	\\
5	&	2	&	1	&	2	&	2	&	-66191/8000	\\
5	&	0	&	1	&	4	&	2	&	27/128	\\
5	&	3	&	1	&	0	&	3	&	14981/(16000 $\sqrt{3}$)	\\
5	&	1	&	1	&	2	&	3	&	9 $\sqrt{3}$/64	\\
5	&	2	&	1	&	0	&	4	&	-66191/16000	\\
5	&	0	&	1	&	2	&	4	&	27/128	\\
5	&	1	&	1	&	0	&	5	&	-9 $\sqrt{3}$/128	\\
5	&	0	&	1	&	0	&	6	&	9/128	\\
5	&	2	&	2	&	2	&	0	&	-189/64	\\
5	&	1	&	2	&	2	&	1	&	15 $\sqrt{3}$/512	\\
5	&	2	&	2	&	0	&	2	&	-189/64	\\
5	&	1	&	2	&	0	&	3	&	-5 $\sqrt{3}$/512	\\
5	&	2	&	3	&	2	&	0	&	228177/5000	\\
5	&	1	&	3	&	2	&	1	&	45 $\sqrt{3}$/16	\\
5	&	2	&	3	&	0	&	2	&	228177/5000	\\
5	&	1	&	3	&	0	&	3	&	-15 $\sqrt{3}$/16	\\
6	&	4	&	0	&	2	&	0	&	5189/188160	\\
6	&	6	&	0	&	2	&	0	&	358873969/831600000	\\
6	&	2	&	0	&	4	&	0	&	7/6144	\\
6	&	4	&	0	&	4	&	0	&	13690483/10080000	\\
6	&	2	&	0	&	6	&	0	&	128837/672000	\\
6	&	0	&	0	&	8	&	0	&	-45/2048	\\
6	&	3	&	0	&	2	&	1	&	$\sqrt{3}$/256	\\
6	&	5	&	0	&	2	&	1	&	1592807/(1680000 $\sqrt{3}$)	\\
6	&	3	&	0	&	4	&	1	&	367253/(448000 $\sqrt{3}$)	\\
6	&	1	&	0	&	6	&	1	&	-81 $\sqrt{3}$/1024	\\
6	&	4	&	0	&	0	&	2	&	5189/188160	\\
6	&	6	&	0	&	0	&	2	&	358873969/831600000	\\
6	&	2	&	0	&	2	&	2	&	7/3072	\\
6	&	4	&	0	&	2	&	2	&	13690483/5040000	\\
6	&	2	&	0	&	4	&	2	&	932087/224000	\\
6	&	0	&	0	&	6	&	2	&	-45/512	\\
6	&	3	&	0	&	0	&	3	&	-1/(256 $\sqrt{3}$)	\\
6	&	5	&	0	&	0	&	3	&	-1592807/(5040000 $\sqrt{3}$)	\\
6	&	3	&	0	&	2	&	3	&	367253/(672000 $\sqrt{3}$)	\\
6	&	1	&	0	&	4	&	3	&	-135 $\sqrt{3}$/1024	\\
6	&	2	&	0	&	0	&	4	&	7/6144	\\
6	&	4	&	0	&	0	&	4	&	13690483/10080000	\\
6	&	2	&	0	&	2	&	4	&	-406663/224000	\\
6	&	0	&	0	&	4	&	4	&	-135/1024	\\
6	&	3	&	0	&	0	&	5	&	-367253/(1344000 $\sqrt{3}$)	\\
6	&	1	&	0	&	2	&	5	&	-27 $\sqrt{3}$/1024	\\
6	&	2	&	0	&	0	&	6	&	396587/672000	\\
6	&	0	&	0	&	2	&	6	&	-45/512	\\
6	&	1	&	0	&	0	&	7	&	27 $\sqrt{3}$/1024	\\
6	&	0	&	0	&	0	&	8	&	-45/2048	\\
6	&	4	&	1	&	2	&	0	&	-404401/392000	\\
		\hline
    \end{tabular}
\end{minipage}\hspace{0.3cm}
\begin{minipage}[t]{.3\columnwidth}
    \begin{tabular}[t]{cccccc} 
        \hline
        $i_1$ & $i_2$ & $i_3$ & $i_4$ & $i_5$ & coeff \\
        \hline
6	&	2	&	1	&	4	&	0	&	199/1024	\\
6	&	3	&	1	&	2	&	1	&	1390787 $\sqrt{3}$/3136000	\\
6	&	1	&	1	&	4	&	1	&	243 $\sqrt{3}$/2048	\\
6	&	4	&	1	&	0	&	2	&	-404401/392000	\\
6	&	2	&	1	&	2	&	2	&	199/512	\\
6	&	3	&	1	&	0	&	3	&	-1390787/(3136000 $\sqrt{3}$)	\\
6	&	1	&	1	&	2	&	3	&	81 $\sqrt{3}$/1024	\\
6	&	2	&	1	&	0	&	4	&	199/1024	\\
6	&	1	&	1	&	0	&	5	&	-81 $\sqrt{3}$/2048	\\
6	&	2	&	2	&	2	&	0	&	-621/1024	\\
6	&	4	&	2	&	2	&	0	&	-34986681/800000	\\
6	&	2	&	2	&	4	&	0	&	-13986039/320000	\\
6	&	0	&	2	&	6	&	0	&	-27/128	\\
6	&	3	&	2	&	2	&	1	&	-850233 $\sqrt{3}$/40000	\\
6	&	1	&	2	&	4	&	1	&	-243 $\sqrt{3}$/128	\\
6	&	2	&	2	&	0	&	2	&	-621/1024	\\
6	&	4	&	2	&	0	&	2	&	-34986681/800000	\\
6	&	2	&	2	&	2	&	2	&	-13986039/160000	\\
6	&	0	&	2	&	4	&	2	&	-81/128	\\
6	&	3	&	2	&	0	&	3	&	283411 $\sqrt{3}$/40000	\\
6	&	1	&	2	&	2	&	3	&	-81 $\sqrt{3}$/64	\\
6	&	2	&	2	&	0	&	4	&	-13986039/320000	\\
6	&	0	&	2	&	2	&	4	&	-81/128	\\
6	&	1	&	2	&	0	&	5	&	81 $\sqrt{3}$/128	\\
6	&	0	&	2	&	0	&	6	&	-27/128	\\
6	&	2	&	3	&	2	&	0	&	-1377/256	\\
6	&	1	&	3	&	2	&	1	&	-765 $\sqrt{3}$/512	\\
6	&	2	&	3	&	0	&	2	&	-1377/256	\\
6	&	1	&	3	&	0	&	3	&	255 $\sqrt{3}$/512	\\
6	&	2	&	4	&	2	&	0	&	1708047/6250	\\
6	&	1	&	4	&	2	&	1	&	-2727 $\sqrt{3}$/256	\\
6	&	2	&	4	&	0	&	2	&	1708047/6250	\\
6	&	1	&	4	&	0	&	3	&	909 $\sqrt{3}$/256	\\
		\hline
    \end{tabular}
\end{minipage}
\label{tab:z6p11s31}
\end{table}
}

\end{document}